\pdfoutput=1
\RequirePackage{ifpdf}
\ifpdf 
\documentclass[pdftex]{sigma}
\else
\documentclass{sigma}
\fi

\numberwithin{equation}{section}

\newtheorem{Theorem}{Theorem}[section]
\newtheorem{Proposition}[Theorem]{Proposition}
{\theoremstyle{definition}
\newtheorem{Definition}[Theorem]{Definition}}

\usepackage{bbm}
\usepackage{pgf,tikz,graphicx}

\definecolor{jred}{rgb}{0.8,0,0}
\definecolor{jgreen}{rgb}{0,0.7,0}
\definecolor{jblue}{rgb}{0,0,0.8}

\usetikzlibrary{arrows,patterns,shapes,positioning,fit}
\tikzstyle{c} =	[coordinate]
\tikzstyle{vc} = [circle, draw=black, line width=1pt, inner sep=0pt, minimum size=2mm]
\tikzstyle{v} =  [circle, draw=black, line width=.2pt, fill=black, inner sep=0pt, minimum size=1.5mm]
\tikzstyle{vb} =[circle, draw=black, line width=.1pt, fill=jred, inner sep=0pt, minimum size=1mm]
\tikzstyle{vh} =[circle, draw=black, line width=.1pt, fill=jblue, inner sep=0pt, minimum size=1.5mm]
\tikzstyle{vs} =[coordinate]
\tikzstyle{e} =	[draw=jred,line width=1pt]
\tikzstyle{eb}= [draw=jgreen,line width=1pt]
\tikzstyle{es}= [dashed, line width=1pt]
\tikzstyle{ev}= [draw=jgreen,line width=1pt]
\tikzstyle{f} = [line width=0.01pt,dotted,fill=blue, fill opacity=.1]
\tikzstyle{cs}= [ellipse, draw=black, line width=.1pt, fill=jred, inner sep=1pt, minimum size=2mm]

\newcommand{\vertexexstranded}{
\begin{tikzpicture}[x=6ex,y=6ex,baseline={([yshift=-.59ex]current bounding box.center)}] 
    \node [vs]	(12)	at (1,.1)	{};
    \node [vs]	(13)	at (1,0)	{};
    \node [vs]	(14)	at (1,-.1)	{};    
    \node [vs]	(21)	at (.1,1)	{};
    \node [vs]	(24)	at (0,1)	{};
    \node [vs]	(23)	at (-.1,1)	{};    
    \node [vs]	(32)	at (-1,.1)	{};
    \node [vs]	(31)	at (-1,0)	{};
    \node [vs]	(34)	at (-1,-.1)	{};    
    \node [vs]	(41)	at (.1,-1)	{};
    \node [vs]	(42)	at (0,-1)	{};
    \node [vs]	(43)	at (-.1,-1)	{};
    \node [vc, fit=(13) (31)] {};
    \path
    \foreach \i/\j in {14/41,21/12,32/23,43/34}{
    (\i) edge [ev,bend right=40] node [vh] {} (\j)
    }
    \foreach \i/\j in {13/42, 31/24}{
    (\i) edge [ev,bend right=60] node [vh] {} (\j)
    };
    \node [cs,fit=(12) (14)] {};
    \node [cs,fit=(21) (23)] {};
    \node [cs,fit=(32) (34)] {};
    \node [cs,fit=(41) (43)] {};
\end{tikzpicture}
}

\newcommand{\vertexexgraph}{
\begin{tikzpicture}[x=6ex,y=6ex,baseline={([yshift=-.59ex]current bounding box.center)}]    
    \node [vb]	(1)	at (1,0)	{};
    \node [vb]	(2)	at (0,1)	{};
    \node [vb]	(3)	at (-1,0)	{};
    \node [vb]	(4)	at (0,-1)	{};
    \node [v, fill=black, minimum size =3mm]   (0) at (0,0)    {};
    \path
    (1) edge [ev] (2)
    (3) edge [ev] (4)
    \foreach \i in {1,2,3,4}{
    (\i) edge [e] (0)
    }
    \foreach \i/\j in {1/4,4/1,2/3,3/2}{
    (\i) edge [ev,bend right=20] node [c] {} (\j)
    };
\end{tikzpicture}
}

\newcommand{\maptadpole}{
\begin{tikzpicture}[x=2ex,y=2ex,baseline={([yshift=-.59ex]current bounding box.center)}]    
    \node [v]	(1)	at (0,0) {};
    \node [c]	(a)	at (-1,0){};
    \node [c]	(b) at (1,0) {};
    \node [c]	(c1)at (0,1) {};
    \path
    (1) edge [e] (a)
    (1) edge [e] (b)
    (1) edge [e, bend right=90] (c1)
    (1) edge [e, bend left=90] (c1);
\end{tikzpicture}
}

\newcommand{\maptadpoledown}{
\begin{tikzpicture}[x=2ex,y=2ex,baseline={([yshift=-.59ex]current bounding box.center)}]    
    \node [v]	(1)	at (0,0) {};
    \node [c]	(a)	at (-1,0){};
    \node [c]	(b) at (1,0) {};
    \node [c]	(c1)at (0,-1) {};
    \path
    (1) edge [e] (a)
    (1) edge [e] (b)
    (1) edge [e, bend right=90] (c1)
    (1) edge [e, bend left=90] (c1);
\end{tikzpicture}
}

\newcommand{\mapsunrise}{
\begin{tikzpicture}[x=2ex,y=2ex,baseline={([yshift=-.59ex]current bounding box.center)}]    
\node [v]	(1)	at (0,0) {};
\node [v]	(2)	at (2,0) {};
\node [c]	(a)	at (-1,0){};
\node [c]	(b) at (3,0) {};
\path
(1) edge [e] (a)
(2) edge [e] (b)
(1) edge [e] (2)
(1) edge [e, bend right=90] (2)
(1) edge [e, bend left=90] (2);
\end{tikzpicture}
}

\newcommand{\mapex}{
\begin{tikzpicture}[x=4ex,y=4ex,baseline={([yshift=-.59ex]current bounding box.center)}]    
    \node [v, label=30:1]	(1)	at (-1.5,0) {};
    \node [v, label=150:2, label=-90:3, label=90:4]	(234)	at (0,0) {};
    \node [v, label=-90:5, label=90:6, label=30:7]	(789)at (1.5,0) {};
    \node [c]	(9)	at (2.25,0) {};
    \path
    (1) edge [e] (234)
    (234) edge [e, bend right=80] (789)
    (234) edge [e, bend left=80] (789)
    (789) edge [e] (9);
\end{tikzpicture}
}

\newcommand{\mapexpq}{
\begin{tikzpicture}[x=3ex,y=3ex,baseline={([yshift=-.59ex]current bounding box.center)}]    
    \node [v]	(1)	at (-1.5,0) {};
    \node [v, label=0:$q$]	(234)	at (0,0) {};
    \node [v, label=30:$p$]	(789)   at (1.5,0) {};
    \node [c]	(9)	at (2.25,0) {};
    \path
    (1) edge [e] (234)
    (234) edge [e, bend right=80] (789)
    (234) edge [e, bend left=80] (789)
    (789) edge [e] (9);
\end{tikzpicture}
}

\newcommand{\mapexvg}{
\begin{tikzpicture}[x=4ex,y=4ex,baseline={([yshift=-.59ex]current bounding box.center)}]        
    \node [v]	(v1)	at (-2.5,0) {};
    \node [c]	(c1)	at (-3,0) 	{};
    \node [v]	(234)	at (0,0) {};
    \node [v]	(789)at (2.6,0) {};
    \node [vb, label=90:1]	(1)	at (-2,0)	{};
    \node [vb, label=90:2]	(2)	at (-1,0)	{};
    \node [vb, label=-90:3]	(3)	at (.6,-1)	{};
    \node [vb, label=90:4]	(4)	at (.6,1)	{};
    \node [vb, label=90:7]	(9)	at (3.6,0)	{};
    \node [vb, label=90:6]	(8)	at (2,1)	{};
    \node [vb, label=-90:5]	(7)	at (2,-1)	{};
    \path
    (v1) edge [e] (1)
    \foreach \i in {2,3,4}{
    (\i) edge [e] (234)
    }
    \foreach \i in {7,8,9}{
    (\i) edge [e] (789)
    }
    (1) edge [ev, bend right=90] (c1)
    (1) edge [ev, bend left=90] (c1)
    \foreach \i/\j in {2/3,3/4,4/2,7/8,8/9,9/7}{
    (\i) edge [ev] (\j)
    }
    \foreach \i/\j in {1/2,3/7,4/8}{
    (\i) edge [es] (\j)
    };
\end{tikzpicture}
}

\newcommand{\rtedge}{
\begin{tikzpicture}[x=2ex,y=2ex,baseline={([yshift=-.59ex]current bounding box.center)}] 
\node [v]	(v)	at (0,0) 	{};
\node [vb]	(1)	at (-1,0)	{};
\node [vb]	(2)	at (1,0)	{};
\path
\foreach \i in {1,2}{
    (\i) edge [e] (v)
    }
(1)	edge [ev,bend left=35]	(2)
(1)	edge [ev,bend right=35]	(2);
\end{tikzpicture}
}

\newcommand{\onevg}{
\begin{tikzpicture}[x=1.2ex,y=1.2ex,baseline={([yshift=-.59ex]current bounding box.center)}] 
    \node [c]	(c1)	at (-3,0) 	{};
    \node [vb]	(1)	at (-2,0)	{};    
    \path
    (1) edge [ev, bend right=90] (c1)
    (1) edge [ev, bend left=90] (c1);
\end{tikzpicture}
}

\newcommand{\threevg}{
\begin{tikzpicture}[x=.7ex,y=.7ex,baseline={([yshift=-.59ex]current bounding box.center)}] 
\node [vb]	(1)	at (-1,0)	{};
\node [vb]	(2)	at (.6,-1)	{};
\node [vb]	(3)	at (.6,1)	{};
\path
\foreach \i/\j in {1/2,2/3,3/1}{
    (\i) edge [ev] (\j)
    };
\end{tikzpicture}
}

\newcommand{\rtvf}{
\begin{tikzpicture}[x=2ex,y=2ex,baseline={([yshift=-.59ex]current bounding box.center)}]
\scriptsize{
\node [v]	(v)	at (0,0) 	{};
\node [vb]	(1)	at (-1,1)	{};
\node [vb]	(2)	at (-1,-1)	{};
\node [vb]	(3)	at (1,-1)	{};
\node [vb]	(4)	at (1,1)	{};
\path
\foreach \i in {1,2,3,4}{
    (\i) edge [e] (v)
    }
(1) edge [ev]  node [c, label=left:$c$] {} (2)
\foreach \i/\j in {2/3,3/4,4/1}{
    (\i) edge [ev] (\j)
    };
}
\end{tikzpicture}
}

\newcommand{\rtvfg}{
\begin{tikzpicture}[x=.7ex,y=.7ex,baseline={([yshift=-.59ex]current bounding box.center)}]
\node [vb]	(1)	at (-1,1)	{};
\node [vb]	(2)	at (-1,-1)	{};
\node [vb]	(3)	at (1,-1)	{};
\node [vb]	(4)	at (1,1)	{};
\path
\foreach \i/\j in {1/2,2/3,3/4,4/1}{
    (\i) edge [ev] (\j)
    };
\end{tikzpicture}
}

\newcommand{\rttadpole}{
\begin{tikzpicture}[x=2ex,y=2ex,baseline={([yshift=-1.2ex]current bounding box.center)}]
\node [v]	(v)	at (0,0) 	{};
\node [vb]	(1)	at (-1,1)	{};
\node [vb]	(2)	at (-1,-1)	{};
\node [vb]	(3)	at (1,-1)	{};
\node [vb]	(4)	at (1,1)	{};
\path
\foreach \i in {1,2,3,4}{
    (\i) edge [e] (v)
    }
\foreach \i/\j in {1/2,2/3,3/4,4/1}{
    (\i) edge [ev] (\j)
    }
(1)	edge [es,bend left=60]	(4);
\end{tikzpicture}
}

\newcommand{\rttadpolec}{
\begin{tikzpicture}[x=2ex,y=2ex,baseline={([yshift=-1.2ex]current bounding box.center)}]
\scriptsize{
\node [v]	(v)	at (0,0) 	{};
\node [vb]	(1)	at (-1,1)	{};
\node [vb]	(2)	at (-1,-1)	{};
\node [vb]	(3)	at (1,-1)	{};
\node [vb]	(4)	at (1,1)	{};
\path
\foreach \i in {1,2,3,4}{
    (\i) edge [e] (v)
    }
(1) edge [ev]  node [c, label=left:$c$] {} (2)
\foreach \i/\j in {2/3,3/4,4/1}{
    (\i) edge [ev] (\j)
    }
(1)	edge [es,bend left=60]	(4);
}
\end{tikzpicture}
}

\newcommand{\rttadpolepq}{
\begin{tikzpicture}[x=2ex,y=2ex,baseline={([yshift=-1.2ex]current bounding box.center)}]
\scriptsize{
\node [v]	(v)	at (0,0) 	{};
\node [vb]	(1)	at (-1,1)	{};
\node [vb]	(2)	at (-1,-1)	{};
\node [vb]	(3)	at (1,-1)	{};
\node [vb]	(4)	at (1,1)	{};
\path
\foreach \i in {1,2,3,4}{
    (\i) edge [e] (v)
    }
(1) edge [ev]  node [c, label=left:$p_1$] {} (2)
(1) edge [ev]  node [c, label=above:$q_2$]{} (4)
\foreach \i/\j in {2/3,3/4,4/1}{
    (\i) edge [ev] (\j)
    };
\draw [es] (1) arc [start angle=180, end angle=0, radius=1];
}
\end{tikzpicture}
}

\newcommand{\rttadpoleb}{
\begin{tikzpicture}[x=2ex,y=2ex,baseline={([yshift=-.59ex]current bounding box.center)}]
\scriptsize{
\node [v]	(v)	at (0,0) 	{};
\node [vb]	(1)	at (-1,1)	{};
\node [vb]	(2)	at (-1,-1)	{};
\node [vb]	(3)	at (1,-1)	{};
\node [vb]	(4)	at (1,1)	{};
\path
\foreach \i in {1,2,3,4}{
    (\i) edge [e] (v)
    }
(1) edge [ev]  node [c, label=above:$c$] {} (4)
\foreach \i/\j in {1/2,2/3,3/4}{
    (\i) edge [ev] (\j)
    }
(2)	edge [es,bend right=60]	(3);
}
\end{tikzpicture}
}

\newcommand{\rttadpolebpq}{
\begin{tikzpicture}[x=2ex,y=2ex,baseline={([yshift=.49ex]current bounding box.center)}]
\scriptsize{
\node [v]	(v)	at (0,0) 	{};
\node [vb]	(1)	at (-1,1)	{};
\node [vb]	(2)	at (-1,-1)	{};
\node [vb]	(3)	at (1,-1)	{};
\node [vb]	(4)	at (1,1)	{};
\path
\foreach \i in {1,2,3,4}{
    (\i) edge [e] (v)
    }
(1) edge [ev]  node [c, label=left:$p_2$] {} (2)
(2) edge [ev]  node [c, label=below:$q_1$] {} (3)
\foreach \i/\j in {3/4,4/1}{
    (\i) edge [ev] (\j)
    };
\draw [es] (2) arc [start angle=180, end angle=360, radius=1];
}
\end{tikzpicture}
}

\newcommand{\rtsunrisec}{
\begin{tikzpicture}[x=2ex,y=2ex,baseline={([yshift=-.6ex]current bounding box.center)}]
\scriptsize{
\begin{scope}[rotate=-45]
\node [v]	(v1)at (0,0) 	{};
\node [vb]	(1)	at (-1,1)	{};
\node [vb]	(2)	at (-1,-1)	{};
\node [vb]	(3)	at (1,-1)	{};
\node [vb]	(4)	at (1,1)	{};
\end{scope}
\begin{scope}[xshift=1.5cm,rotate=-45]
\node [v]	(v2)at (0,0) 	{};
\node [vb]	(5)	at (-1,1)	{};
\node [vb]	(6)	at (-1,-1)	{};
\node [vb]	(7)	at (1,-1)	{};
\node [vb]	(8)	at (1,1)	{};
\end{scope}
\path
\foreach \i in {1,2,3,4}{
    (\i) edge [e] (v1)
    }
\foreach \i in {5,6,7,8}{
    (\i) edge [e] (v2)
    }
(1) edge [ev]  node [c, label=135:$c$] {} (2)
(8) edge [ev]  node [c, label=45:$c$] {} (5)
\foreach \i/\j in {2/3,3/4,4/1,5/6,6/7,7/8}{
    (\i) edge [ev] (\j)
    }
\foreach \i/\j in {1/5,3/7,4/6}{
    (\i) edge [es] (\j)
    };
}
\end{tikzpicture}
}

\newcommand{\rtsunrisepq}{
\begin{tikzpicture}[x=2ex,y=2ex,baseline={([yshift=-.6ex]current bounding box.center)}]
\scriptsize{
\begin{scope}[rotate=-45]
\node [v]	(v1)at (0,0) 	{};
\node [vb]	(1)	at (-1,1)	{};
\node [vb]	(2)	at (-1,-1)	{};
\node [vb]	(3)	at (1,-1)	{};
\node [vb]	(4)	at (1,1)	{};
\end{scope}
\begin{scope}[xshift=1.5cm,rotate=-45]
\node [v]	(v2)at (0,0) 	{};
\node [vb]	(5)	at (-1,1)	{};
\node [vb]	(6)	at (-1,-1)	{};
\node [vb]	(7)	at (1,-1)	{};
\node [vb]	(8)	at (1,1)	{};
\end{scope}
\path
\foreach \i in {1,2,3,4}{
    (\i) edge [e] (v1)
    }
\foreach \i in {5,6,7,8}{
    (\i) edge [e] (v2)
    }
(1) edge [ev]  node [c, label=135:$p_1$] {} (2)
(2) edge [ev]  node [c, label=225:$p_2$] {} (3)
(3) edge [ev]  node [c, label=right:$q_1$] {} (4)
(4) edge [ev]  node [c, label=right:$q_2$] {} (1)
\foreach \i/\j in {5/6,6/7,7/8,8/5}{
    (\i) edge [ev] (\j)
    }
\foreach \i/\j in {1/5,3/7,4/6}{
    (\i) edge [es] (\j)
    };
}
\end{tikzpicture}
}

\newcommand{\rtsunrisesub}[1]{
\begin{tikzpicture}[x=2ex,y=2ex,baseline={([yshift=-.6ex]current bounding box.center)}]
\scriptsize{
\begin{scope}[rotate=-45]
\node [v]	(v1)at (0,0) 	{};
\node [vb]	(1)	at (-1,1)	{};
\node [vb]	(2)	at (-1,-1)	{};
\node [vb]	(3)	at (1,-1)	{};
\node [vb]	(4)	at (1,1)	{};
\end{scope}
\begin{scope}[xshift=1.5cm,rotate=-45]
\node [v]	(v2)at (0,0) 	{};
\node [vb]	(5)	at (-1,1)	{};
\node [vb]	(6)	at (-1,-1)	{};
\node [vb]	(7)	at (1,-1)	{};
\node [vb]	(8)	at (1,1)	{};
\end{scope}
\path
\foreach \i in {1,2,3,4}{
    (\i) edge [e] (v1)
    }
\foreach \i in {5,6,7,8}{
    (\i) edge [e] (v2)
    }
(1) edge [ev]  node [c, label=135:$c$] {} (2)
(8) edge [ev]  node [c, label=45:$c$] {} (5)
\foreach \i/\j in {2/3,3/4,4/1,5/6,6/7,7/8}{
    (\i) edge [ev] (\j)
    }
\foreach \i/\j in {#1}{
    (\i) edge [es] (\j)
    };
}
\end{tikzpicture}
}

\newcommand{\rtfisha}{
\begin{tikzpicture}[x=2ex,y=2ex,baseline={([yshift=-1.2ex]current bounding box.center)}]
\scriptsize{
\begin{scope}[rotate=-90]
\node [v]	(v1)at (0,0) 	{};
\node [vb]	(1)	at (-1,1)	{};
\node [vb]	(2)	at (-1,-1)	{};
\node [vb]	(3)	at (1,-1)	{};
\node [vb]	(4)	at (1,1)	{};
\end{scope}
\begin{scope}[xshift=1.3cm]
\node [v]	(v2)at (0,0) 	{};
\node [vb]	(5)	at (-1,1)	{};
\node [vb]	(6)	at (-1,-1)	{};
\node [vb]	(7)	at (1,-1)	{};
\node [vb]	(8)	at (1,1)	{};
\end{scope}
\path
\foreach \i in {1,2,3,4}{
    (\i) edge [e] (v1)
    }
\foreach \i in {5,6,7,8}{
    (\i) edge [e] (v2)
    }
(1) edge [ev]  node [c, label=above:$c$] {} (2)
(8) edge [ev]  node [c, label=above:$c$] {} (5)
\foreach \i/\j in {2/3,3/4,4/1,5/6,6/7,7/8}{
    (\i) edge [ev] (\j)
    }
\foreach \i/\j in {1/5,4/6}{
    (\i) edge [es] (\j)
    };
}
\end{tikzpicture}
}

\newcommand{\rtfishpq}[3]{
\begin{tikzpicture}[x=2ex,y=2ex,baseline={([yshift=-.59ex]current bounding box.center)}]
\scriptsize{
\begin{scope}[rotate=-90]
\node [v]	(v1)at (0,0) 	{};
\node [vb]	(1)	at (-1,1)	{};
\node [vb]	(2)	at (-1,-1)	{};
\node [vb]	(3)	at (1,-1)	{};
\node [vb]	(4)	at (1,1)	{};
\end{scope}
\begin{scope}[xshift=1.3cm]
\node [v]	(v2)at (0,0) 	{};
\node [vb]	(5)	at (-1,1)	{};
\node [vb]	(6)	at (-1,-1)	{};
\node [vb]	(7)	at (1,-1)	{};
\node [vb]	(8)	at (1,1)	{};
\end{scope}
\path
\foreach \i in {1,2,3,4}{
    (\i) edge [e] (v1)
    }
\foreach \i in {5,6,7,8}{
    (\i) edge [e] (v2)
    }
(1) edge [ev]  node [c, label=above:$#1$] {} (2)
(3) edge [ev]  node [c, label=below:$#3$] {} (4)
(1) edge [ev]  node [c, label=right:$#2$] {} (4)
\foreach \i/\j in {2/3,3/4,4/1,5/6,6/7,7/8,8/5}{
    (\i) edge [ev] (\j)
    }
\foreach \i/\j in {1/5,4/6}{
    (\i) edge [es] (\j)
    };
}
\end{tikzpicture}
}

\newcommand{\rtfishb}{
\begin{tikzpicture}[x=2ex,y=2ex,baseline={([yshift=-.59ex]current bounding box.center)}]
\scriptsize{
\node [v]	(v1)at (0,0) 	{};
\node [vb]	(1)	at (-1,1)	{};
\node [vb]	(2)	at (-1,-1)	{};
\node [vb]	(3)	at (1,-1)	{};
\node [vb]	(4)	at (1,1)	{};
\begin{scope}[xshift=1.3cm]
\node [v]	(v2)at (0,0) 	{};
\node [vb]	(5)	at (-1,1)	{};
\node [vb]	(6)	at (-1,-1)	{};
\node [vb]	(7)	at (1,-1)	{};
\node [vb]	(8)	at (1,1)	{};
\end{scope}
\path
\foreach \i in {1,2,3,4}{
    (\i) edge [e] (v1)
    }
\foreach \i in {5,6,7,8}{
    (\i) edge [e] (v2)
    }
(1) edge [ev]  node [c, label=left:$c$] {} (2)
(5) edge [ev]  node [c, label=left:$c$] {} (6)
\foreach \i/\j in {2/3,3/4,4/1,6/7,7/8,8/5}{
    (\i) edge [ev] (\j)
    }
\foreach \i/\j in {4/5,3/6}{
    (\i) edge [es] (\j)
    };
}
\end{tikzpicture}
}

\newcommand{\rhvfg}[1]{
\begin{tikzpicture}[rotate=-90,x=.7ex,y=.7ex,baseline={([yshift=-.59ex]current bounding box.center)}] 
\scriptsize{
\node [vb]	(b1)	at (-1,1)	{};
\node [vb]	(w1)	at (-1,-1)	{};
\node [vb]	(b2)	at (1,-1)	{};
\node [vb]	(w2)	at (1,1)	{};
\path
\foreach \i/\j in {1/2}{
   	(w\i) edge [ev]   (b\j)
	(b\i) edge [ev]  node [c, label=right:$#1$] {} (w\j)
	}
\foreach \i in {1,2}{
	(w\i)	edge [ev,bend left=30]	(b\i)
	(w\i)	edge [ev,bend right=30]	(b\i)
	};
}
\end{tikzpicture}
}

\newcommand{\rhtadpolea}{
\begin{tikzpicture}[rotate=90, x=2ex,y=2ex,baseline={([yshift=-1.2ex]current bounding box.center)}] 
\scriptsize{
\node [v]	(1)	    at (0,0)	{};
\node [vb]	(b1)	at (-1,1)	{};
\node [vb]	(w1)	at (-1,-1)	{};
\node [vb]	(b2)	at (1,-1)	{};
\node [vb]	(w2)	at (1,1)	{};
\path
\foreach \i/\j in {1/2}{
   	(w\i) edge [ev]   (b\j)
	(b\i) edge [ev]  node [c, label=left:$c$] {} (w\j)
	}
\foreach \i in {1,2}{
	(w\i)	edge [ev,bend left=30]	(b\i)
	(w\i)	edge [ev,bend right=30]	(b\i)
	}
\foreach \i in {w1,b1,w2,b2}{
   	(\i) edge [e] (1)
	}
(w2)	edge [es,bend left=90]	(b2);
}
\end{tikzpicture}
}

\newcommand{\rhtadpoleapq}{
\begin{tikzpicture}[rotate=90, x=2ex,y=2ex,baseline={([yshift=-1.2ex]current bounding box.center)}] 
\scriptsize{
\node [v]	(1)	    at (0,0)	{};
\node [vb]	(b1)	at (-1,1)	{};
\node [vb]	(w1)	at (-1,-1)	{};
\node [vb]	(b2)	at (1,-1)	{};
\node [vb]	(w2)	at (1,1)	{};
\path
\foreach \i/\j in {1/2}{
   	(w\i) edge [ev]   (b\j)
	(b\i) edge [ev]  node [c, label=left:$p_1$] {} (w\j)
	}
\foreach \i in {1,2}{
	(w\i)	edge [ev,bend left=30]	(b\i)
	(w\i)	edge [ev,bend right=30]	(b\i)
	}
\foreach \i in {w1,b1,w2,b2}{
   	(\i) edge [e] (1)
	}
(w2)	edge [es,bend left=90]	(b2);
}
\end{tikzpicture}
}

\newcommand{\rhtadpoleb}{
\begin{tikzpicture}[x=2ex,y=2ex,baseline={([yshift=-.3ex]current bounding box.center)}] 
\scriptsize{
\node [v]	(1)	    at (0,0)	{};
\node [vb]	(b1)	at (-1,1)	{};
\node [vb]	(w1)	at (-1,-1)	{};
\node [vb]	(b2)	at (1,-1)	{};
\node [vb]	(w2)	at (1,1)	{};
\path
\foreach \i/\j in {1/2}{
   	(w\i) edge [ev]  node [c, label=below:$c$] {} (b\j)
	(b\i) edge [ev]   (w\j)
	}
\foreach \i in {1,2}{
	(w\i)	edge [ev,bend left=30]	(b\i)
	(w\i)	edge [ev,bend right=30]	(b\i)
	}
\foreach \i in {w1,b1,w2,b2}{
   	(\i) edge [e] (1)
	}
(b1)	edge [es,bend left=60]	(w2);
}
\end{tikzpicture}
}

\newcommand{\rhtadpolebpq}{
\begin{tikzpicture}[x=2ex,y=2ex,baseline={([yshift=-1.2ex]current bounding box.center)}] 
\scriptsize{
\node [v]	(1)	    at (0,0)	{};
\node [vb]	(b1)	at (-1,1)	{};
\node [vb]	(w1)	at (-1,-1)	{};
\node [vb]	(b2)	at (1,-1)	{};
\node [vb]	(w2)	at (1,1)	{};
\path
\foreach \i/\j in {1/2}{
   	(w\i) edge [ev]  (b\j)
	(b\i) edge [ev]  node [c, label=above:$q_1$] {} (w\j)
	}
(w1)	edge [ev,bend left=30] node [c, label=left:{$p_2,p_3$}] {}	(b1)
(w1)	edge [ev,bend right=30]	(b1)
\foreach \i in {2}{
	(w\i)	edge [ev,bend left=30]	(b\i)
	(w\i)	edge [ev,bend right=30]	(b\i)
	}
\foreach \i in {w1,b1,w2,b2}{
   	(\i) edge [e] (1)
	};
\draw [es] (b1) arc [start angle=180, end angle=0, radius=1];
}
\end{tikzpicture}
}

\newcommand{\rhsunrise}[5]{
\begin{tikzpicture}[x=2ex,y=2ex,baseline={([yshift=-.6ex]current bounding box.center)}]
\scriptsize{
\begin{scope}[rotate=-45]
\node [v]	(v1)at (0,0) 	{};
\node [vb]	(1)	at (-1,1)	{};
\node [vb]	(2)	at (-1,-1)	{};
\node [vb]	(3)	at (1,-1)	{};
\node [vb]	(4)	at (1,1)	{};
\end{scope}
\begin{scope}[xshift=1.5cm,rotate=-45]
\node [v]	(v2)at (0,0) 	{};
\node [vb]	(5)	at (-1,1)	{};
\node [vb]	(6)	at (-1,-1)	{};
\node [vb]	(7)	at (1,-1)	{};
\node [vb]	(8)	at (1,1)	{};
\end{scope}
\path
\foreach \i in {1,2,3,4}{
    (\i) edge [e] (v1)
    }
\foreach \i in {5,6,7,8}{
    (\i) edge [e] (v2)
    }
(1) edge [ev]  node [c, label=135:$#1$] {} (2)
(2) edge [ev, bend left=30]  node [c, label=225:$#2$] {} (3)
(2) edge [ev, bend right=30] (3)
(4) edge [ev, bend right=30]  node [c, label=right:$#3$] {} (1)
(4) edge [ev, bend left=30]   (1)
(3) edge [ev]  node [c, label=right:$#4$] {} (4)
(5) edge [ev]  node [c, label=45:$#5$] {} (8)
\foreach \i/\j in {6/7,8/5}{
    (\i) edge [ev] (\j)
    }
\foreach \i\j in {5/6,7/8}{
	(\i)	edge [ev,bend left=30]	(\j)
	(\i)	edge [ev,bend right=30]	(\j)
	}
\foreach \i/\j in {1/5,3/7,4/6}{
    (\i) edge [es] (\j)
    };
}
\end{tikzpicture}
}

\newcommand{\rhfish}[4]{
\begin{tikzpicture}[x=2ex,y=2ex,baseline={([yshift=-.59ex]current bounding box.center)}]
\scriptsize{
\begin{scope}[rotate=-90]
\node [v]	(v1)at (0,0) 	{};
\node [vb]	(1)	at (-1,1)	{};
\node [vb]	(2)	at (-1,-1)	{};
\node [vb]	(3)	at (1,-1)	{};
\node [vb]	(4)	at (1,1)	{};
\end{scope}
\begin{scope}[xshift=1.8cm]
\node [v]	(v2)at (0,0) 	{};
\node [vb]	(5)	at (-1,1)	{};
\node [vb]	(6)	at (-1,-1)	{};
\node [vb]	(7)	at (1,-1)	{};
\node [vb]	(8)	at (1,1)	{};
\end{scope}
\path
\foreach \i in {1,2,3,4}{
    (\i) edge [e] (v1)
    }
\foreach \i in {5,6,7,8}{
    (\i) edge [e] (v2)
    }
(1) edge [ev]  node [c, label=above:$#1$] {} (2)
(1) edge [ev,bend left=30]  node [c, label=right:$#2$] {} (4)
(1) edge [ev,bend right=30] (4)
(3) edge [ev]  node [c, label=below:$#3$] {} (4)
(5) edge [ev]  node [c, label=above:$#4$] {} (8)
\foreach \i/\j in {3/4,6/7,8/5}{
    (\i) edge [ev] (\j)
    }
\foreach \i\j in {2/3,5/6,7/8}{
	(\i)	edge [ev,bend left=30]	(\j)
	(\i)	edge [ev,bend right=30]	(\j)
	}
\foreach \i/\j in {1/5,4/6}{
    (\i) edge [es] (\j)
    };
}
\end{tikzpicture}
}

\newcommand{\boxd}[1]{\boxed{\phantom{\Big(}#1\phantom{\Big)}}}

\newcommand{\cdef}[1]{\emph{#1}}
\def\d{\mathrm{d}}

\newcommand{\N}{\mathbb N}

\newcommand{\Q}{\mathbb Q}
\newcommand{\R}{\mathbb R}

\newcommand{\In}{\subset}
\newcommand{\id}{\mathrm{id}}

\newcommand{\og}{\gamma}
\newcommand{\tg}{\Gamma}

\newcommand{\sg}{\Theta}

\newcommand{\V}{\mathcal{V}}

\newcommand{\E}{\mathcal{E}}

\newcommand{\Fx}{\mathcal{F}^{\mathrm{ext}}}
\newcommand{\Fxi}{\widetilde{\mathcal{F}}^{\mathrm{ext}}}
\newcommand{\Fi}{\mathcal{F}^{\mathrm{int}}}

\newcommand{\nv}{V}

\newcommand{\nei}{E}

\newcommand{\nf}{F}

\newcommand{\nc}{K}

\newcommand{\es}{{\sigma_2}}

\newcommand{\OG}{\mathbf{G}_{1}}

\newcommand{\sgr}{\In}

\newcommand{\br}{\partial}			 

\newcommand{\uca}{\mathcal{A}}
\newcommand{\btg}{\mathcal{G}}

\newcommand{\hfd}{\mathcal{H}^{\textrm{f2g}}}

\newcommand{\one}{\mathbbm 1}
\newcommand{\cop}{\Delta}
\newcommand{\cou}{\epsilon}
\newcommand{\conp}{*}
\newcommand{\anti}{S}

\newcommand{\amp}{A}
\newcommand{\ramp}{A_\textsc{r}}
\newcommand{\rota}{R}
\newcommand{\ranti}{S^\textsc{a}_\textsc{r}}
\newcommand{\w}{\omega}
\newcommand{\sdd}{\omega^{\textrm{sd}}}
\newcommand{\gdeg}{\omega^\textsc{g}}
\newcommand{\uvd}{d_{\rk}}
\newcommand{\ks}{{2\zeta}} 

\newcommand{\pa}{\tilde{p}_1}
\newcommand{\pb}{\tilde{p}_2}
\newcommand{\pc}{\tilde{p}_i}
\newcommand{\pd}{\tilde{p}_{23}}

\newcommand{\std}{D}				
\newcommand{\rk}{r} 	
\newcommand{\gd}{d} 
\newcommand{\gf}{\Phi}
\newcommand{\Sk}{S_0} 	
\newcommand{\Sia}{S_\textsc{ia}} 		

\newcommand{\vp}{\pmb{p}}

\begin{document}
\allowdisplaybreaks

\newcommand{\arXivNumber}{2103.01136}

\renewcommand{\thefootnote}{}

\renewcommand{\PaperNumber}{094}

\FirstPageHeading

\ShortArticleName{Renormalization in Combinatorially Non-Local Field Theories}

\ArticleName{Renormalization in Combinatorially Non-Local\\ Field Theories: the BPHZ Momentum Scheme\footnote{This paper is a~contribution to the Special Issue on Algebraic Structures in Perturbative Quantum Field Theory in honor of Dirk Kreimer for his 60th birthday. The~full collection is available at \href{https://www.emis.de/journals/SIGMA/Kreimer.html}{https://www.emis.de/journals/SIGMA/Kreimer.html}}}

\Author{Johannes TH\"URIGEN~$^{\rm ab}$}
\AuthorNameForHeading{J.~Th\"urigen}

\Address{$^{\rm a)}$~Mathematisches Institut, Westf\"alische Wilhelms-Universit\"at M\"unster,\\
\hphantom{$^{\rm a)}$}~Einsteinstr.~62, 48149 M\"unster, Germany}
\EmailD{\href{mailto:johannes.thuerigen@uni-muenster.de}{johannes.thuerigen@uni-muenster.de}}
\URLaddressD{\url{http://www.uni-muenster.de/mathphys/en/u/thuerigen/}}

\Address{$^{\rm b)}$~Institut f\"ur Physik/Mathematik, Humboldt-Universit\"at zu Berlin,\\
\hphantom{$^{\rm b)}$}~Unter den Linden 6, 10099 Berlin, Germany}

\ArticleDates{Received February 28, 2021, in final form October 24, 2021; Published online October 27, 2021}

\Abstract{Various combinatorially non-local field theories are known to be renormalizable. Still, explicit calculations of amplitudes are very rare and restricted to matrix field theory. In this contribution I want to demonstrate how the BPHZ momentum scheme in terms of the Connes--Kreimer Hopf algebra applies to any combinatorially non-local field theory which is renormalizable. This algebraic method improves the understanding of known results in noncommutative field theory in its matrix formulation. Furthermore, I use it to provide new explicit perturbative calculations of amplitudes in tensorial field theories of rank $r>2$.}

\Keywords{non-local field theory; renormalization; Hopf algebras; multiple polylogarithms}

\Classification{05C10; 16T05; 16T30; 81T15; 81T18; 81T32}

\begin{flushright}
\begin{minipage}{60mm}
\it Dedicated to Dirk Kreimer\\ on the occasion of his 60th birthday
\end{minipage}
\end{flushright}

\renewcommand{\thefootnote}{\arabic{footnote}}
\setcounter{footnote}{0}

\section{Introduction}

Since Dirk Kreimer's seminal work \cite{Kreimer:1998te} it is well known that a Hopf algebra of divergent Feynman diagrams is the universal structure underlying renormalization in perturbative quantum field theory \cite{Connes:2000km,Connes:2001hk,Kreimer:2002br}.
A classical example is BPHZ momentum subtraction as used for the general proof of renormalization \cite{Bogoliubow:1957eu,Hepp:1966bx,Zimmermann:1969up}.
In \cite{Thurigen:2021vy} I have shown that the Connes--Kreimer Hopf algebra describes renormalization not only in field theories with point-like interactions but also with full generality in combinatorially non-local field theories (cNLFT) such as non-commutative and matrix field theory \cite{Grosse:2004bm,Hock:20,Wulkenhaar:2019}, tensorial field theories \cite{BenGeloun:2014gp, BenGeloun:2013fw} and group field theory \cite{Carrozza:2013uq,Freidel:2005jy, Oriti:2003uw}.

So far, explicit calculations of renormalized amplitudes in cNLFT depending on external kinematic variables are only known for noncommutative field theory and its matrix-field representation using the BPHZ momentum scheme \cite{Blaschke:2013kq,Grosse:2017ep,Hock:20}.
In fact, exact solutions for matrix field theory have been found recently \cite{Branahl:2020vc,Grosse:2020uo, Panzer:2018tg} and the perturbative results are mainly used as a check of consistency \cite{Grosse:2017ep,Hock:20}.
For other examples of cNLFT, renormalizability has been proven for various field theories with tensorial interactions \cite{BenGeloun:2014gp,BenGeloun:2013fw, BenGeloun:2013dl,Carrozza:2013uq, Carrozza:2014bh,Carrozza:2014ee} but no explicit results of amplitudes, let alone of full renormalized correlation functions, are known.\footnote{Exact solutions similar to matrix field theory \cite{Panzer:2018tg} have been found for a super-renormalizable rank-3 tensor field in \cite{Pascalie:2019tu}; due to the restriction to a single tensorial interaction and renormalization even of convergent amplitudes, a closer look reveals that this is still the mentioned matrix theory in disguise.} Using the Hopf-algebraic structure of renormalization here I will give new results for explicit amplitudes of perturbative tensorial field theory.

Explicit perturbative calculations in matrix field theory use a version of Zimmermann's forest formula \cite{Zimmermann:1969up} adapted to ribbon graphs \cite{Grosse:2017ep}.
While agreement with exact non-perturbative solutions shows that this implementation of the BPHZ momentum scheme is perfectly valid, it uses a very peculiar definition of subdiagrams which are not themselves ribbon graphs, thus strictly speaking in a different class of diagrams.
Here I will show how BPHZ momentum subtraction can be defined on any cNLFT in Kreimer's way using the Hopf algebra of divergent diagrams and a subtraction operator as a Taylor expansion
which is a Rota--Baxter operator on the algebra of amplitudes.
This gives two results:
First, it reproduces in a very natural way the known perturbative calculations in matrix field theory \cite{Grosse:2017ep,Hock:20}.
Second, it provides the algorithm to calculate renormalized amplitudes in any cNLFT in a straightforward way.
I will demonstrate this for the example of tadpole and sunrise diagrams in field theory with tensor-invariant interactions.

The physical meaning of amplitudes in cNLFT is not as obvious as for scattering events of high energy particles described by standard-model quantum field theories.
Matrix field theory can be understood as a representation of field theory of elementary particles on non-commutative (Moyal-deformed) spacetime which still allows for such scattering interpretation.
On the other hand, one main application of theories with tensor fields of higher rank $\rk>2$ are models of quantum gravity as their perturbative series is a sum over $\rk$-dimensional combinatorial (pseudo) manifolds \cite{Gurau:2011dw} and one can build models which have gravitational amplitudes on such discrete manifolds as lattice gauge theories \cite{Carrozza:2013uq,Freidel:2005jy, Oriti:2003uw}, known as spin foam models \cite{Perez:2013uz}.
Then, the main challenge is to recover continuous $\std=4$ dimensional spacetime in some critical regime of the theory \cite{Oriti:2007qd, Rivasseau:2012jk} in the sense of continuous random geometries \cite{Aldous:1991kr,Lionni:2019vc,Marckert:2006fs};
to this end, non-perturbative results like \cite{Pithis:2020gy} are necessary.
However, also perturbation theory might be the right starting point towards this goal, either by identification of appropriate linear combinations of amplitudes in the perturbative series related to non-perturbative structures as found for example via topological recursion \cite{Branahl:2020vca, Branahl:2020vc} or exploiting the Hopf-algebraic structure of perturbative renormalization leading to Kreimer's combinatorial Dyson--Schwinger equations \cite{Bergbauer:2006vm, Broadhurst:2001hw, Kreimer:2006gg, Kreimer:2006ft}.
Thus, the explicit perturbative results presented here might also be seen as a starting point to investigate the non-perturbative regime of cNLFT.

\section{Feynman rules in cNLFT}

In the following I introduce the general structure of renormalization as captured by a Hopf algebra \cite{Kreimer:1998te,Kreimer:2002br}.
In parallel, I explain its application to combinatorially non-local field theory.
This is largely based on~\cite{Thurigen:2021vy} where one can find the precise details of combinatorial and algebraic de\-finitions and statements.
Here, the emphasis is on the practical use for perturbative calculations in field theory.

A perturbative expansion in coupling constants is a formal power series labelled by Feynman diagrams.
Feynman rules associate to each diagram $\tg$ an amplitude $\amp_\tg$, i.e., the coefficient in this power series.
This can be understood as a map from diagrams to amplitudes
\begin{gather*}
\amp \colon \ \btg \to \uca, \qquad \tg \mapsto \amp(\tg) = \amp_\tg .
\end{gather*}
In the perturbative series these amplitudes are summed and
for renormalization it will also be necessary to multiply them which demands an algebra $\uca$ of amplitudes.
This structure is mirrored in a $\Q$-algebra $\btg$ generated by the set of diagrams in which multiplication is simply the disjoint union of diagrams, $m_\btg(\tg_1\otimes\tg_2):=\tg_1\sqcup\tg_2$.
Both algebras are unital with unit $u_\btg\colon \mathbb{Q}\to\btg,q\mapsto q\one$ where $\one\in\btg$ is the empty diagram and $u_\uca$ maps to the amplitude $A(\one)=1$.
The map $A$ must then be an algebra homomorphism.

A combinatorially non-local field theory (cNLFT) is characterized by a pairing of arguments at interaction vertices which leads to strand graphs as Feynman diagrams (called \cdef{$2$-graphs} in~\cite{Thurigen:2021vy}\footnote{I thank an unknown referee for pointing out the possible confusions with the notion of a ``two-graph'' as defined by G.~Higman \cite{Seidel:1973ww, Taylor:1971uv}. For this reason I will here use ``strand graph'' for what is defined as ``2-graph'' in~\cite{Thurigen:2021vy}.}).
That is, order-$n$ interactions are of the form
\begin{gather}\label{eq:nonlocalinteraction}
\Sia[\gf] = \lambda_{\og}\int\prod_{i=1}^n \d\vp_i \prod_{(ia,jb)}\delta\big(p_i^a-p_j^b\big) \prod_{i=1}^n {\gf}(\vp_i),
\end{gather}
where the field $\gf$ depends on $\rk$ arguments $\vp=\big(p^1,\dots ,p^\rk\big)$ of which each $p^a$ is in a $\gd$-dimensional manifold. For simplicity I will consider $p^a\in\R^\gd$ throughout this paper.
The product over pairs $(ia,jb)$ means that for each argument $p_i^a$ there is a convolution in terms of the Dirac distribution $\delta\big(p_i^a-p_j^b\big)$ with exactly one other argument~$p_j^b$.
Diagrammatically, these convolutions cannot be fully captured by a vertex in a graph but it is necessary to add a second layer called strands.
For example,
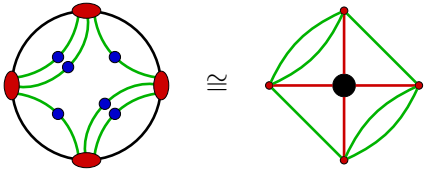
\begin{gather*}
\vertexexstranded \quad \cong \quad \vertexexgraph
\end{gather*}
is the diagrammatic representation of an interaction vertex of $n=4$ fields (red) with $\rk=3$ arguments each (green) which are convoluted pairwise (blue dots); this is equivalent to, on the right-hand side of the equation,
a vertex (black) with half-edges (red) and additional edges (green) adjacent to these half-edges which characterize the interaction vertex by a graph.
Thus, interactions are labelled in general by \cdef{vertex graphs} $\og$ and so are their coupling constants $\lambda_\og$.
A~set of graphs $\OG^v$ determines the diagrammatics of a given theory.

Feynman diagrams follow as usual by Wick contraction of the fields according to a quadratic Gaussian part of the action.
The arguments $p^a$ of the field $\gf$ are not necessarily ordered, i.e., the quadratic part
might pair the $\rk$ arguments according to an arbitrary permutation~$\es$,\footnote{Note that for a given permutation this quadratic form might not be positive which, however, is necessary to yield a Gaussian measure. With the sum $\sum_{\es}$ over an appropriate set of such permutations $\es$ the quadratic form becomes positive and thus Gaussian. I thank one of the referees for pointing this out.}
\begin{gather}\label{eq:kineticaction}
\Sk(\gf) = \frac{1}{2}\int \d\vp_i\d\vp_j \gf(\vp_i) \sum_{\es} {\prod_{a=1}^\rk \delta\big(p_i^a-p_j^{\es(a)}\big)}
\bigg({\sum_{a=1}^\rk |p_i^a|^\ks + \mu}\bigg) \gf(\vp_j) ,
\end{gather}
where $\mu$ is a mass parameter and the scaling $\ks$ of the propagator is in general parametrized by a $\zeta>0$, usually $\zeta\in{}]0,1]$.
This leads to the general class of strand graphs defined in \cite{Thurigen:2021vy} under the name of ``2-graphs'' and based on $\cite{Oriti:2015kv,Thurigen:2015}$.
Still, in many theories this permutation is fixed. For example, in Hermitian matrix field theory \cite{Hock:20, Wulkenhaar:2019}, the strand graphs are combinatorial maps (ribbon graphs ) and $\es$ is fixed by the orientation of the vertices, e.g., in
\begin{gather}\label{eq:mapexample}
\tg= \mapex \cong \mapex \cong \mapexvg
\end{gather}
the pairing of strands adjacent to half edges $i\in\{1,2,3,4,5,6,7\}$ along the edges $\{1,2\}$, $\{3,5\}$ and $\{4,6\}$ is determined by the orientation of the vertices $(1)$, $(234)$ and $(576)$ (see Fig.~2 in~\cite{Thurigen:2021vy} for further details).
Following a strand through the strand graph defines a face which is an internal face if the strand is closed and external otherwise.
In the example $\tg$ there is one internal face $(3564)$ and one external face~$(764212357)$.
The external structure of a strand graph~$\tg$ is the vertex graph~$\og$ obtained by deleting all internal edges and internal faces which is the boundary $\og=\br\tg$.

The above Gaussian and interaction parts of the action, equations~\eqref{eq:kineticaction} and \eqref{eq:nonlocalinteraction}, lead to an identification of variables $p^a$ along strands and thus to the following Feynman rules:
For a~strand graph $\tg$, the amputated amplitude $A_\tg$ is an integral consisting of
\begin{enumerate}\itemsep=0pt
 \item[1)] a Lebesgue integral $\int_{\R^\gd} \d q_f$ for each internal face $f\in\Fi_\tg$, integrating over the variable identifying $q_f=p_i^a$ for all strands $a$ at half edges $i$ along $f$,
 \item[2)] a propagator factor $\frac{1}{\sum_{a=1}^\rk |p_i^a|^\ks + \mu}$ for each internal edge $e=\{i,j\}\in\E_\tg$,
 \item[3)] a coupling constant $\lambda_{\og_v}$ for each vertex $v\in\V_\tg$ with vertex graph $\og=\og_v$ and
\end{enumerate}
These rules define the algebra homomorphism $\amp\colon \btg\to\uca$ to the algebra $\uca$ of (formal) integral functions with rational functions as integrands,
\begin{gather*}
\amp_\tg \colon \ \{p_f\}_{f\in\Fxi_\tg}
\mapsto \amp_\tg(\{p_f\}):=
\prod_{v\in \V_\tg} \lambda_{\og_v}
\prod_{f\in \Fi_\tg}\int_{\R^\gd} \d q_f
\prod_{\{i,j\}\in \E_\tg} \frac{1}{\sum_{a=1}^\rk |p_i^a|^\ks + \mu} .
\end{gather*}
They depend on external variables $p_f$ for each external face $f\in\Fxi_\tg$ adjacent to \emph{at least one internal edge}, given again by identifying $p_f=p^a_i$ for all half edges $i$ and adjacent strands~$a$ along~$f$.
Importantly, only faces in this subset $\Fxi\In\Fx$ of external faces play a role: If a face is not adjacent to any internal edge, there is no propagator in which the corresponding variable occurs and the amputated amplitude $\amp_\tg$ does not depend on it. This is in contrast to local field theory where loop amplitudes always depend on all external variables.

Multiplication in the algebra $\uca$ of amplitudes is given by multiplication of integrands.
That is, the product of two amplitudes $A,B\in\uca$ with
\begin{gather*}
A(p_1,\dots ,p_m)=c_A \int\prod_i\d q_i I_A(p_1,\dots ,p_m,q_1,\dots ,q_k)
\end{gather*} and
\begin{gather*} B(r_1,\dots ,r_n)=c_B \int\prod_j\d s_j I_B(r_1,\dots ,r_n,s_1,\dots ,s_l),
\end{gather*} where the integrands $I_A$ and $I_B$ are rational functions is
\begin{gather*} 
(A\cdot B)(p_a,r_b) := c_A c_B \int \prod_{i=1}^k \d q_i \prod_{j=1}^l \d s_j I_A(p_1,\dots ,p_m,q_1,\dots ,q_k) I_B(r_1,\dots ,r_n,s_1,\dots ,s_l) .
\end{gather*}
This is relevant since some external variables $p_a$ of the amplitude $A$ might coincide with some integration variables $s_j$ of $B$, i.e., $p_a=s_j$ for some $a\in\{1,\dots ,n\}$ and $j\in\{1,\dots ,l\}$, and vice versa.
Accordingly, these are not external variables of the product $A\cdot B$.

To give an example of an amplitude, for $\tg$ in equation~\eqref{eq:mapexample} there is one external variable $p$ for the external face and one integration variable $q$ for the internal face leading to the amputated amplitude
\begin{gather*}
\amp_{\tg}(p) = \amp\left(\mapexpq \right)(p) =
\lambda_{\onevg} \lambda_{\threevg}^2 \int_{\R^\gd} \d q \frac{1}{2|p|^\ks+\mu}\frac{1}{\big(|p|^\ks+|q|^\ks+\mu\big)^2} .
\end{gather*}
For $d\ge\ks$ this integral does not converge.
To render the algebra of formal amplitudes $\uca$ an algebra of well defined functions it is necessary to introduce a regularization cutoff $\Lambda$, that is consider the algebra $\uca^\Lambda$ of the same integrals but
integrating all internal variables over the domain $|q|^\ks\le\Lambda^\ks$,
\begin{gather}\label{eq:regularizedamplitude}
\amp^\Lambda_\tg(\{p_f\}):=
\prod_{v\in \V_\tg} \lambda_{\og_v}
\prod_{f\in \Fi}\int_\Lambda \d q_f
\prod_{\{i,j\}\in \E} \frac{1}{\sum_{a=1}^\rk |p_i^a|^\ks + \mu} .
\end{gather}
For a well-defined limit $\Lambda\to\infty$ renormalization is needed.

\section[BPHZ renormalization using the Hopf algebra of strand graphs]{BPHZ renormalization using the Hopf algebra\\ of strand graphs}

To renormalize a divergent amplitude one subtracts a counter term which exactly eliminates the divergent part and leaves a finite result.
To this end one needs a measure of how much a divergent amplitude is divergent.
A useful starting point is the power $\w_\tg$ of the asymptotic scaling in the cutoff $\Lambda$ of the regularized amplitude \eqref{eq:regularizedamplitude}, i.e., $\amp^\Lambda_\tg\sim\Lambda^{\w_\tg}$.
A simple way to determine this quantity is to count the number of integrals and propagators in the integration variables in the amplitude.
For \eqref{eq:regularizedamplitude} this gives the \cdef{superficial degree of divergence}
\begin{gather}\label{eq:superficialdegreeofdivergence}
\sdd(\tg) = \gd \nf_\tg - \ks \nei_\tg + \sum_{v\in\V_\tg} \w(\og_v),
\end{gather}
where $\nf_\tg=|\Fi_\tg|$ is the number of internal faces, $\nei_\tg=|\E_\tg|$ is the number of (internal) edges and a scaling $\lambda_\og \sim \Lambda^{\w(\og)}$ of the couplings with vertex graph $\og=\og_v$ for each vertex $v\in\V_\tg$ in the strand graph $\tg$ is assumed in addition.
A strand graph $\tg$ is \cdef{superficially divergent} iff~$\sdd(\tg)\ge0$.
These properties are specific to a given combinatorially non-local field theory~$T$ which is defined by a set of interactions given by its vertex graphs $\OG^v$ together with their scaling weights $\w\colon \OG^v\to\R$, the propagators with scaling~$\ks$ and the dimension~$d$.

The theorem of Bogoliubov and Parasiuk \cite{Bogoliubow:1957eu} with proof completed by Hepp \cite{Hepp:1966bx} and further improvements by Zimmermann leading to the forest formula
\cite{Zimmermann:1969up} states that in a renormalizable local quantum field theory all divergences can be renormalized by counter terms for each superficially divergent one-particle irreducible (1PI) diagram.
Kreimer \cite{Kreimer:1998te} has further significantly improved the understanding of the combinatorics of the forest formula (in particular, and renormalization in much more general) showing that it has an algebraic structure given by a~Hopf algebra of divergent Feynman diagrams.
In \cite{Thurigen:2021vy} I have shown that this Hopf algebra exists not only in field theories with point-like interactions but also for theories with any combinatorially non-local interactions,
in this way generalizing and improving similar statements for some specific examples \cite{Raasakka:2013wa,Tanasa:2013kh, Tanasa:2007xa}.
Here I demonstrate how this works explicitly using the BPHZ momentum scheme.

The challenge to define the right counter term to renormalize a divergent amplitude lies in the nested structure of divergent subdiagrams.
In cNLFT, just like in local quantum field theory~\cite{Borinsky:2018}, a strand graph $\sg$ is a \cdef{subgraph} $\sg\sgr\tg$ of a strand graph $\tg$ iff it has a subset of edges $\E_\sg\In\E_\tg$, including the adjacent strands, and the same structure otherwise~\cite{Thurigen:2021vy}.
A \cdef{primitively divergent} strand graph is then one which has no proper subgraphs which are superficially divergent.
For a primitively divergent strand graph $\tg$ one defines the renormalized amplitude $\ramp(\tg)$ directly by a subtraction operation which is a linear operator $\rota\colon \uca\to\uca$ as
\begin{gather}\label{eq:countertermprimitive}
\ramp\colon \ \btg\to \ramp(\tg) := (\amp - \rota\circ\amp)(\tg) .
\end{gather}
The explicit form of $\rota$ is exactly what specifies a particular renormalization scheme.
For concreteness, let us here define $\rota$ for the BPHZ momentum scheme of cNLFT:
\begin{Definition}[BPHZ momentum scheme]\label{def:BPHZ}
The operator $\rota\colon \uca^\Lambda\to\uca^\Lambda$ on the unital commutative algebra $\uca^\Lambda$ of functions of the form $\amp^\Lambda_\tg$, equation~\eqref{eq:regularizedamplitude}, with degree of divergence $\w\big(\amp^\Lambda_\tg\big):=\sdd(\tg)$, equation~\eqref{eq:superficialdegreeofdivergence}, is given by the multivariate Taylor expansion $T^\w$ of order $\w=\w\big(\amp^\Lambda_\tg\big)$,
\begin{gather}\label{eq:Taylorsubtraction}
\rota\big[\amp^\Lambda_\tg\big](\{p_f\}):= \big(T^\w
_{\{p_f\}} \amp^\Lambda_\tg\big)(\{p_f\})
= \sum_{|\vec{k}|\le\w(\amp^\Lambda_\tg)} \frac{1}{\vec{k}!} \frac{\partial^{|\vec{k}|} \amp_\tg^\Lambda}{\prod_f \partial p_f^{k_f}} \big(\{p_f=0\}\big) \prod_{f\in\Fxi_\tg} p_f^{k_f} ,
\end{gather}
where the sum is over multi-indices $\vec{k}$ of length $|\vec{k}|=\sum_{f\in\Fxi_\tg} k_f$.
Note that the underlying strand-graph structure is only used to encode the form of the function $\amp(\{p_f\})=\amp^\Lambda_\tg(\{p_f\})$
but the definition of $\rota$ is otherwise independent of it (any $\Lambda$-regularized integral has a degree $\w$ given by its scaling $\Lambda^\w$).
\end{Definition}

For divergent diagrams $\tg$ with subdivergences the whole counter term is a nested sum over counter terms for the various superficially divergent subgraphs according to Zimmermann's forest formula~\cite{Zimmermann:1969up}.
These terms are products of counter terms of subdiagrams~$\sg\sgr\tg$ and amplitudes of remaining diagrams in which the subdiagram is shrunken to a vertex, that is contractions~$\tg/\sg$.
For strand graphs $\sg\sgr\tg$ the definition of such contraction $\tg/\sg$ is that there is a vertex for each connected component of $\sg$ and all internal edges and internal strands (and thus internal faces) of $\sg$ are deleted (see \cite[Definition~3.2]{Thurigen:2021vy} for further detail).
This defines a coproduct $\cop$ on the algebra of superficially divergent strand graphs $\hfd_T$ of a given renormalizable theory~$T$ as
\begin{gather*}
\cop\colon \ \hfd_T \to \hfd_T\otimes\hfd_T, \qquad \cop\tg:=\sum_{\sg\sgr\tg, \sg\in\hfd_T} \sg \otimes \tg/\sg .
\end{gather*}
For this it is crucial that $\hfd_T$ is closed under contractions which is satisfied due to the property of renormalizability $\w(\br\tg)=\sdd(\tg)-\delta_\tg$ where the $\tg$-dependent part $\delta_\tg$ needs to satisfy $\delta_\tg = \delta_\sg+\delta_{\tg/\sg}$ for all subgraphs~$\sg\sgr\tg$ \cite{Thurigen:2021vy}.
Together with the counit $\cou\colon \hfd_T \to \Q$ defined as the projector to strand graphs without any edges, $\hfd$ is a coalgebra and due to compatibility with multiplication also a bialgebra \cite[Proposition~4.2]{Thurigen:2021vy}.

To implement the forest formula, the crucial object is an inverse to the algebra homomorphism of amplitudes $\amp\colon \hfd\to\uca$
with respect to the convolution product, i.e.,
\begin{gather*}
\phi \conp \psi := m_{\uca} \circ (\phi\otimes\psi) \circ \cop
\end{gather*}
for arbitrary such homomorphisms $\phi, \psi\colon \hfd\to\uca$.
Such inverse is a natural consequence if~$\hfd$ is a Hopf algebra, i.e., has a \cdef{coinverse}, also called \cdef{antipode}, $\anti\colon \hfd\to\hfd$ defined by $\anti \conp \id = \id \conp \anti = u \circ \cou$.
In fact, one can show that the bialgebra $\hfd_T$ of superficially divergent strand graphs in $T$ is a Hopf algebra \cite[Theorem~5.1]{Thurigen:2021vy} using the possibility to express~$\anti$ recursively as
\begin{gather*}
\anti = -(\anti\conp P),
\end{gather*}
where $P=\id-u\circ\cou$ is the projector to the augmentation ideal, i.e., all strand graphs in $\hfd_T$ with at least one edge \cite{Borinsky:2018, Manchon:2004vi}. This works because $\hfd$ is graded with respect to the number of edges.
As a direct consequence, the inverse of an algebra homomorphism $\phi$ is $\anti^\phi=\anti\circ\phi$.
Both facts can now be used to recursively define the counter term
\begin{align}
\ranti\colon \ & \hfd\to\uca,\label{eq:counterterm}\\
& \ranti(\tg_0)=1, \qquad \ranti(\tg)
= -\rota\big[\big(\ranti\conp \amp\circ P\big)(\tg)\big]= -
\sum_{\sg\subsetneq\tg,\,\sg\in\hfd_T} \rota\big[ \ranti(\sg)\amp(\tg/\sg)\big]\nonumber
\end{align}
for any strand graph $\tg_0$ without edges, and any $\tg$ in the augmentation ideal.
Then, the renormalized amplitude is simply
\begin{gather*}
\boxd{
\ramp := \ranti \conp \amp .
}
\end{gather*}
The counter term map $\ranti$ is an algebra homomorphism, in particular $\ranti(\tg_1\tg_2)=\ranti(\tg_1)\ranti(\tg_2)$, if the subtraction map $\rota$ is a Rota--Baxter operator, that is satisfies
\begin{gather*}
\rota[AB]+R[A]R[B]=R[R[A]B+AR[B]] .
\end{gather*}

\begin{Proposition}
The BPHZ momentum subtraction operator $\rota=T^\w$ $($Definition~{\rm \ref{def:BPHZ})} is a~Rota--Baxter operator.
\end{Proposition}

\begin{proof}The statement is equivalent to Proposition~9.1 in~\cite{EbrahimiFard:2007fb}.
There, the subtraction operator is defined on the integrands of the amplitudes which are ignorant about the degree $\w$ with the consequence that each specific~$R$ depends on $\w$ and it defines therefore a Rota--Baxter family (which still is very close to a Rota--Baxter operator by Proposition~9.2 therein).
As $R$ is defined in Theorem~\ref{def:BPHZ} via the scaling $\Lambda^\w$ of $\Lambda$-regularized integrals,
this construction is not necessary.\footnote{I thank one of the referees for pointing out this subtlety.}
\end{proof}

\section[BPHZ renormalization in tensorial Phi4{2,2} and Phi4{1,3} theory]{BPHZ renormalization in tensorial $\boldsymbol{\gf^4_{2,2}}$ and $\boldsymbol{\gf^4_{1,3}}$ theory}

To demonstrate how BPHZ renormalization works, I give the example of tadpole and sunrise amplitudes in the just renormalizable complex tensorial field theories of rank $\rk=2$ and $\rk=3$.
A~tensorial $\gf^n_{\gd,\rk}$ theory is a cNLFT of a scalar field $\gf$ with $\rk$ arguments in a $\gd$-dimensional manifold with tensor-invariant interactions of maximal order $n$ (degree of the vertex) \cite{BenGeloun:2014gp}.
An interaction is tensor-invariant if its vertex graph $\og$ is $\rk$-regular and edge-coloured, i.e., each vertex in~$\og$ is adjacent to exactly $\rk$ edges with distinct labels $c=1,2,\dots ,\rk$ \cite{Bonzom:2012bg}.
This corresponds to fixing the position of each argument in the field and pairing only arguments in the same position.
As a consequence, also the strand graphs $\tg$ labelling the perturbative expansion can be represented as coloured graphs,\footnote{Strictly speaking, this is only true for theories without multi-trace vertices, that is vertices whose vertex graphs have more than one connected component. Otherwise, the full coloured strand-graph structure is necessary~\cite{Thurigen:2021vy}.}
now with $\rk+1$ colours.
In the following I will consider a~complex field~$\gf$, just for the reason that the vertex graphs then have to be furthermore bipartite which simplifies the theory space further.
Thus, the action of such theory is
\begin{gather*}
S[\gf,\bar{\gf}] = \left(\prod_{c=1}^\rk \int_{\R^\gd}\d p^c\right) \bar{\gf}(\vp)\left(\sum_{c=1}^\rk |p^c|^\ks + \mu \right)\gf(\vp)
+ \sum_{\og\in\OG^v} \lambda_\og
\operatorname{Tr}_\og\big(\gf,\bar{\gf}\big),
\end{gather*}
where $\operatorname{Tr}_\og$ denotes the convolution of the field, detailed in equation~\eqref{eq:nonlocalinteraction}, according to a given graph $\og$ in the set $\OG^v$ of the theory's interaction vertex graphs.

With weight $\w(\og_v) = \uvd - {d^v}(\uvd-\ks)/2$ for each $d^v$-valent vertex with vertex graph $\og_v$, the superficial degree of divergence in tensorial $\gf^n_{\gd,\rk}$ theory is \cite{BenGeloun:2014gp, Thurigen:2021vy}
\begin{gather*}
\sdd(\tg) = \uvd - \frac{\uvd-\ks}{2}\nv_{\br\tg}
-\gd\left(\frac{2 \gdeg_\tg-2\gdeg_{\br\tg}}{(\rk-1)!} + \nc_{\br\tg}-1\right),
\end{gather*}
where $\uvd=\gd(\rk-1)$ is the equivalent of dimension compared to local field theory, $\nc_{\br\tg}$ is the number of connected components of the boundary graph and $\gdeg\in\N$ is the Gurau degree of a~coloured graph~\cite{Gurau:2012ek,Gurau:2016}.
It is a quantity generalizing the genus~$g$ in the $\rk=2$ case, though not a~topological invariant.
Crucially, the overall Gurau degree is conserved under the coproduct~\cite{Raasakka:2013wa} such that renormalizability is given as $\sdd(\tg)=\w(\br\tg)-\delta_\tg$ with $\delta_\tg = \gd\frac{2 \gdeg_\tg-2\gdeg_{\br\tg}}{(\rk-1)!}$.
One can further show that $\gdeg_\tg\ge\gdeg_{\br\tg}$~\cite{BenGeloun:2013fw}
such that the whole term in brackets is always greater or equal to zero and renormalizable theories are determined by the first two terms.
Thus, a tensorial $\gf^n_{\gd,\rk}$ theory is just renormalizable if $n=\frac{2\uvd}{\uvd-\ks}$.

Just-renormalizable quartic theories are those with $\uvd=\gd(\rk-1)=4\zeta$.
The tensorial equivalent to standard local $\phi^4_4$ theory with quadratic propagator $\zeta=1$ are therefore theories with $\uvd=4$, that is theories with $(\gd,\rk)=(4,2),(2,3)$ or $(1,5)$. Let us here simplify explicit renormalization even further and chose $\zeta=1/2$. The two just renormalizable theories are then tensorial~$\gf^4_{2,2}$ and~$\gf^4_{1,3}$ theory.
The divergence degree in these two theories is
\begin{gather*}
\sdd(\tg)= \gd \nf - \nei = 2 - \frac{1}{2} \nv_{\br\tg} -
\begin{cases}
2\left(2 g_\tg + \nc_{\br\tg}-1\right) & \textrm{for $\gf^4_{2,2}$ theory}, \\
\left(\gdeg_\tg - g_{\br\tg} + \nc_{\br\tg}-1\right) & \textrm{for $\gf^4_{1,3}$ theory}.
\end{cases}
\end{gather*}
In both cases there is only a single type of marginal quartic interaction given by the graph $\rtvfg$ and $\rhvfg{c}$ for $c=1,2,3$ respectively.
It is not necessary to include the quartic double-trace interaction.

For these two cases I will present the explicit BPHZ-renormalized amplitudes $\amp_\tg$ for the tadpole and sunrise diagrams.
Complex $\gf^4_{2,2}$ theory is very similar to the quartic Hermitian matrix field theory known as the Grosse-Wulkenhaar model or quartic Kontsevich model \cite{Grosse:2004bm,Hock:20,Wulkenhaar:2019}, i.e., has the same power counting and the same set of divergent diagrams with identical amplitudes;
for this theory I simply reproduce known results~\cite{Hock:20} with the Hopf algebra method.
The rank $\rk=3$ tensorial theory is one of the first for which renormalizability was proven~\cite{BenGeloun:2013dl}.
In this case the calculations presented here are new explicit results of amplitudes in tensorial field theory.

\subsubsection*{The tadpole in $\boldsymbol{\gf^4_{2,2}}$ theory}

The only primitive 2-point amplitude in $\gf^4_{2,2}$ theory is the one for the tadpole strand graph.
There are two versions,
$\maptadpole\cong\rttadpolec$ with colour $c=1$ and
$\maptadpoledown\cong\rttadpolec$ with $c=2$ when using the convention that in the representation via combinatorial maps $\maptadpole$ and $\maptadpoledown$ one has always the external variable $p_1$ associated to the face of colour $c=1$ on the upper face and~$p_2$ respectively on the lower face.
Its divergence degree is $\sdd(\maptadpole)=\sdd(\maptadpoledown)=2-1=1$.
There are no subdivergences and thus by definition~\eqref{eq:regularizedamplitude}
\begin{align*}
\ramp\bigg(\rttadpole\bigg) &= \big(\ranti\conp\amp\big)\bigg(\rttadpole\bigg) \nonumber\\
&= m_\uca\big(\ranti\otimes\amp\big)\cop\bigg(\rttadpole\bigg)
= m_\uca(\ranti\otimes\amp)\bigg(\rtvf\otimes\rttadpole +\rttadpole\otimes\rtedge\bigg) \nonumber\\
&= \amp\bigg(\rttadpole\bigg) + \ranti\bigg(\rttadpole\bigg)\amp\left(\rtedge\right)
= \amp\bigg(\rttadpole\bigg) -\rota\Big[\amp\bigg(\rttadpole\bigg)\Big]
\end{align*}
since $\amp\left(\rtedge\right)=1$ as there are no edges in a single vertex.
This calculation works in general for any primitive diagram and thus validates equation~\eqref{eq:countertermprimitive}.
Explicit calculation in the BPHZ momentum scheme according to Definition~\ref{def:BPHZ} gives then
\begin{align}
\ramp\bigg(\rttadpole\bigg)(p_1)&\equiv\ramp\bigg(\rttadpolepq\bigg)
= \lambda_{\rtvfg}\left(1-T^1_{p_1}\right)\int_{\R^2}\d q_2 \frac{1}{|p_1|+|q_2|+\mu} \nonumber\\
&= \lambda_{\rtvfg}\int_0^\infty 2\pi |q_2|\d |q_2| \left(
\frac{1}{|p_1|+|q_2|+\mu} - \frac{1}{|q_2|+\mu} + \frac{|p_1|}{(|q_2|+\mu)^2}
\right) \nonumber\\
&= 2\pi \mu\lambda_{\rtvfg} \big( (\pa+1 )\log (\pa+1 ) - \pa \big)\label{eq:matrixtadpole}
\end{align}
with $\pa={|p_1|}/{\mu}$. The tadpole amplitude for colour $c=2$ gives the same with $p_2$ instead of~$p_1$.
This result is exactly the same as in Hermitian matrix field theory~\cite{Hock:20}.

\subsubsection*{Tadpole diagrams in $\boldsymbol{\gf^4_{1,3}}$ theory}

In $\gf^4_{1,3}$ theory, the tadpole has not only versions for each colour but also two possibilities where to have the edge. In particular, the two differ in degree of divergence (using $\sdd(\tg)=\nf-\nei$),
\begin{gather*}
\sdd\bigg(\rhtadpolea \bigg) = 2-1 = 1 \qquad \textrm{and} \qquad
\sdd\bigg(\rhtadpoleb \bigg) = 1-1 = 0 .
\end{gather*}
The renormalized amplitude of the first one for $c=1$ is
\begin{align*}
&\ramp\bigg(\rhtadpoleapq \bigg) = \lambda_{\rhvfg{1}} \big(1-T^1_{p_1}\big)\int_{\R}\d q_2 \int_{\R}\d q_3 \frac{1}{|p_1|+|q_2|+ |q_3|+\mu} \nonumber\\
&= \lambda_{\rhvfg{1}} \int_0^\infty 2 \d q_2 \int_0^\infty 2 \d q_3 \left(
\frac{1}{|p_1|+q_2+q_3+\mu} - \frac{1}{q_2+q_3+\mu} + \frac{|p_1|}{(q_2+q_3+\mu)^2}
\right) \nonumber\\
&= 4 \mu\lambda_{\rhvfg{1}} \big(\left(\pa+1\right)\log\left(\pa+1\right) - \pa \big)
\end{align*}
with $\pa={|p_1|}/{\mu}$. Up to a constant it is the same as in the $\rk=2$ matrix case, equation~\eqref{eq:matrixtadpole}. This is not surprising as the external variable is the same and the two internal faces with one-dimensional integration variables $q_2$ and $q_3$ behaves effectively like one face with a two-dimensional variable.

The first amplitude specific to a higher rank $\rk=3$ is 
\begin{align}
&\ramp\bigg(\rhtadpolebpq \bigg) = \lambda_{\rhvfg{1}} \big(1-T^0_{p_2,p_3}\big) \int_{\R}\d q_1 \frac{1}{|q_1|+|p_2|+ |p_3|+\mu} \nonumber\\
&= \lambda_{\rhvfg{1}} \int_0^\infty 2 \d q_1 \left(
\frac{1}{q_1+|p_2|+|p_3|+\mu} - \frac{1}{q_1+\mu}
\right)
= -2 \lambda_{\rhvfg{1}} \log(\pd+1)\label{eq:amplituderank3tadpole2}
\end{align}
with $\pd=(|p_2|+|p_3|)/\mu$. In the full 1PI 2-point Green's function one has a sum over the different colours $c=1,2,3$ for which in general the couplings $\lambda_{\rhvfg{c}}$ are distinct.

\subsubsection*{The sunrise diagram in $\boldsymbol{\gf^4_{2,2}}$ theory}

The strand-graph version of the sunrise diagram is the first example which shows the Hopf-algebraic formulation of Zimmermann's forest formula in its full beauty. In complex $\gf^4_{2,2}$ theory the sunrise combinatorial map corresponds to the colouring
\begin{gather*}
\mapsunrise\cong\rtsunrisec
\qquad \textrm{and has} \quad
\sdd\Big(\mapsunrise \Big) = 2\cdot2 - 3 = 1 .
\end{gather*}
(The other choice of colouring of this strand graph corresponds to a non-planar map of cylinder topology which is not divergent, $\sdd=0-3=-3$.)
The sunrise strand graph has three 1PI subgraphs of which only two are divergent,
\begin{gather}
\sdd\bigg(\rtfisha\bigg)=
\sdd\bigg(\rtfishb\bigg)=2\cdot1-2=0,\nonumber\\
\sdd\bigg(\rtsunrisesub{1/5,3/7}\bigg)=0-2=-2 .\label{eq:sunrisesubgraphs}
\end{gather}
As they do not have any divergent 1PI subgraphs, they are primitives.
Accordingly, the coproduct in $\hfd$ is
\begin{align*}
\cop\bigg(\rtsunrisec\bigg) ={} &
\rtvf \rtvf \otimes \rtsunrisec
+\rtfisha \otimes \rttadpoleb \nonumber\\
&{}+\rtfishb \otimes \rttadpolec
+\rtsunrisec \otimes \rtedge
\end{align*}
which yields the renormalized amplitude
\begin{align*}
\ramp\bigg(\rtsunrisepq\bigg) ={}& \amp\bigg(\rtsunrisepq\bigg)
+\ranti\bigg(\rtfishpq{p_1}{q_2}{q_1}\bigg)\amp\bigg(\rttadpolebpq \bigg)\\
&{}+\ranti\bigg(\rtfishpq{q_2}{q_1}{p_2}\bigg)\amp\bigg(\rttadpolepq \bigg)
+\ranti\bigg(\rtsunrisepq\bigg) . \nonumber
\end{align*}
In the last term, $\ranti$ acts on a divergent diagram which is not primitive such that according to~\eqref{eq:counterterm} there is a nontrivial sum
\begin{align}
\ranti\bigg(\rtsunrisepq\bigg) = -\rota\Bigg[ &
\amp\bigg(\rtsunrisepq\bigg)
-\rota\bigg[\amp\bigg(\rtfishpq{p_1}{q_2}{q_1}\bigg)\bigg]\amp\bigg(\rttadpolebpq \bigg) \nonumber\\
&{}-\rota\bigg[\amp\bigg(\rtfishpq{q_2}{q_1}{p_2}\bigg)\bigg]\amp\bigg(\rttadpolepq \bigg)
\Bigg].\label{eq:countertermexpansion}
\end{align}
Putting everything together and applying $\rota=T^\w$ one arrives at the finite integral
\begin{gather*}
\ramp(p_1,p_2)=\lambda^2_{\rtvfg} \big(1-T^1_{p_1,p_2}\big) \int_{\R^2} \d q_1 \int_{\R^2} \d q_2 \bigg(
\frac{1}{|p_1|+|q_2|+\mu}\frac{1}{|q_1|+|q_2|+\mu}\frac{1}{|q_1|+|p_2|+\mu} \nonumber\\
\hphantom{\ramp(p_1,p_2)=\lambda^2_{\rtvfg} \big(1-T^1_{p_1,p_2}\big)}{}
+\frac{1}{|q_1|+|p_2|+\mu}\big({-}T^0_{p_1,q_1}\big)\frac{1}{|p_1|+|q_2|+\mu}\frac{1}{|q_1|+|q_2|+\mu} \nonumber\\
\hphantom{\ramp(p_1,p_2)=\lambda^2_{\rtvfg} \big(1-T^1_{p_1,p_2}\big)}{}
+\frac{1}{|p_1|+|q_1|+\mu}\big({-}T^0_{q_2,p_2}\big)\frac{1}{|q_1|+|q_2|+\mu}\frac{1}{|q_2|+|p_2|+\mu}
\bigg) \nonumber\\
\hphantom{\ramp(p_1,p_2)}{} = \lambda^2_{\rtvfg}\frac{4\pi^2 \mu}{\pa+\pb+1} \bigg[
\pa\pb\zeta_2
+(\pa+\pb+1)\sum_{i=1,2}\Big((\pc+1)\log(\pc+1)-\pc\Big) \nonumber\\
\hphantom{\ramp(p_1,p_2)= \lambda^2_{\rtvfg}\frac{4\pi^2 \mu}{\pa+\pb+1} \bigg[}{}
-\prod_{i=1,2}(\pc+1)\log(\pc+1)
+\sum_{i=1,2}\pc(\pc+1)\textrm{Li}_2(-\pc)
\bigg]
\end{gather*}
involving the polylogarithm $\textrm{Li}_2$ and the zeta value $\zeta_2=\pi^2/6$ and again $\pc=|p_i|/\mu$.
In general, any kind of multiple polylogarithms are to be expected due to the form of the integral.
The result reproduces equation~(D.9) in~\cite{Hock:20}.
As the diagram is symmetric under colour change, the amplitude is symmetric under exchange of $p_1$ and $p_2$, i.e., $\ramp(p_2,p_1)=\ramp(p_1,p_2)$.

\subsubsection*{The sunrise diagram in $\boldsymbol{\gf^4_{1,3}}$ theory}

Also in $\gf^4_{1,3}$ there is only one divergent sunrise strand graph out of four possibilities to join two quartic vertices by three edges (without tadpoles).
This diagram has only a logarithmic divergence,
\begin{gather*}
\sdd\bigg(\rhsunrise{c}{}{}{}{c}\bigg) = 3 - 3 = 0 .
\end{gather*}
Out of the three 1PI subgraphs, in analogy to \eqref{eq:sunrisesubgraphs},
only one is divergent which is
\begin{gather*}
\sdd\bigg(\rhfish{c}{}{}{c}\bigg) = 2 - 2 = 0 .
\end{gather*}
Thus, in rank $\rk=3$ there is one counter term less than in the $\rk=2$ case,
\begin{align*}
&\ramp(p_1,p_2,p_3)=\ramp\bigg(\rhsunrise{p_1}{{p_2, p_3}}{{q_2, q_3}}{q_1}{}\bigg) =\amp\bigg(\rhsunrise{p_1}{{p_2, p_3}}{{q_2, q_3}}{q_1}{}\bigg) \\
& \hphantom{\ramp(p_1,p_2,p_3)=}{}
+\ranti\bigg(\rhfish{p_1}{{q_2,q_3}}{q_1}{}\bigg)\amp\bigg(\rhtadpoleapq \bigg)
+\ranti\bigg(\rhsunrise{p_1}{{p_2, p_3}}{{q_2, q_3}}{q_1}{}\bigg) .
\end{align*}
Expanding these counter terms in $\rota$ expressions analogously to~\eqref{eq:countertermexpansion}
and using the explicit subtraction operation $\rota=T^\omega$, this amplitude is
\begin{gather}
\ramp(p_1,p_2,p_3)
=\lambda^2_{\rhvfg{1}} \big(1-T^0_{p_1,p_2,p_3}\big) \int_{\R} \d q_1 \frac{1}{|q_1|+|p_2|+|p_3|+\mu} \nonumber\\
\hphantom{\ramp(p_1,p_2,p_3)=}{}
 \times \big(1-T^0_{p_1,q_1}\big) \int_{\R} \d q_2 \int_{\R} \d q_3
\frac{1}{|q_1|+|q_2|+|q_3|+\mu}\frac{1}{|p_1|+|q_2|+|q_3|+\mu} \nonumber\\
\hphantom{\ramp(p_1,p_2,p_3)}{}
= \lambda^2_{\rhvfg{1}}\frac{6}{1+\pa+\pd} \bigg[
\log(\pd+1)\Big((\pa+1)\log(\pa+1)-\pa-\pd-1\Big)\nonumber\\
\hphantom{\ramp(p_1,p_2,p_3)=}{}
-\log(\pa)(\pa+1)\log(\pa+1)
+\pd\Big(\textrm{Li}_2\left(-\frac1{\pd}\right)+\frac{1}{2}\log^2(\pd)+2\zeta_2\Big) \nonumber\\
\hphantom{\ramp(p_1,p_2,p_3)=}{}
 +(\pa+1)\Big(\textrm{Li}_2\left(\frac1{\pa+1}\right)+\frac{1}{2}\log^2(\pa+1)-\zeta_2\Big)
\bigg],\label{eq:amplituderank3sunrise}
\end{gather}
where again $\pa=|p_1|/\mu$ and $\pd=(|p_2|+|p_3|)/\mu$.
Note that the Taylor operator $T^0_{p_1,p_2,p_3}$ in the first line acts on the whole expression which is following.
One might use shuffle relations of multiple polylogarithms to further simplify the terms of weight two in $\pa+1$ and~$\pd$.

Interestingly, the calculation of the sunrise diagram in the $\rk=3$ theory is simpler than in the $\rk=2$ case due to the lack of overlapping divergences.
This is an instance of the well known breaking of the combinatorial symmetry of matrix field theory which
xleads to asymptotic freedom in quartic tensorial field theories~\cite{Rivasseau:2015im}.
In general, the expectation is that amplitudes in tensorial field theory can always be expanded in multiple polylogarithms for which efficient algorithms exist for computation~\cite{Panzer:2015bu}.

\section{Outlook}

The main purpose of this contribution has been to demonstrate how the Connes--Kreimer Hopf algebra \cite{Connes:2000km,Connes:2001hk,Kreimer:1998te,Kreimer:2002br} can be used for explicit calculations in cNLFT.
Feynman diagrams in cNLFT are strand graphs generalizing the usual Feynman graphs of point-like interactions by adding a second layer of strands along which the degrees of freedom in cNLFT propagate.
The crucial difference to standard quantum field theory is that the external structure of such diagrams is not just given by the number of legs of a given field type but is captured by graphs.
Nevertheless, with the right definitions, the Hopf-algebraic structure relying on the usual coproduct of subgraph contraction is exactly the same as for local quantum field theory \cite{Thurigen:2021vy}.
Thus, it directly gives a convenient algorithm to renormalize Green's functions in cNLFT, such as non-commutative field theory and matrix field theory \cite{Grosse:2004bm,Hock:20,Wulkenhaar:2019} as well as higher rank tensorial and group field theories \cite{BenGeloun:2014gp,BenGeloun:2013fw,Carrozza:2013uq,Freidel:2005jy, Oriti:2003uw}.

While the Hopf algebra unveils the structure of renormalization independent of a specific scheme, here I have demonstrated how it works for BPHZ momentum subtraction \cite{Bogoliubow:1957eu,Hepp:1966bx,Zimmermann:1969up} in two simple examples of tensorial field theory on $\big(\R^d\big)^{\otimes\rk}$.
As a first example, in $\gf^4_{2,2}$ theory, i.e., complex matrix field theory on a 2-dimensional configuration space, the tadpole and sunrise diagram have the same amplitude as in the Grosse--Wulkenhaar model \cite{Grosse:2004bm,Wulkenhaar:2019} and the Hopf-algebraic calculations reproduce the known results \cite{Hock:20}.
Second, as a first example of a field theory of higher rank $\rk>2$, I have calculated the amplitudes for the same diagrams in $\gf^4_{1,3}$ theory.
As to be expected, these calculations give first hints that also in tensorial field theories the algebra of amplitudes is spanned by multiple polylogarithms.
A deeper understanding of this space of amplitudes for tensorial theories would be an interesting research topic in its own.

The renormalized amplitudes of the tensorial versions of the tadpole and sunrise diagrams give examples showing that the UV regime of tensorial field theories \cite{BenGeloun:2014gp} is richer than the large-$N$ limit of the corresponding tensor models \cite{Gurau:2012ek, Gurau:2016}.
The latter is completely described by so-called melonic diagrams, i.e., diagrams which have vanishing Gurau degree $\gdeg=0$.
There have been considerable efforts to find random geometries beyond the melonic regime, for example by enhancement of subleading classes of diagrams \cite{Delepouve:2015hc, Lionni:2019vc, Lionni:2017tk}.
In tensorial field theories, melonic diagrams are the most divergent but not the only divergent ones.
In fact, the tadpole and sunrise amplitudes~\eqref{eq:amplituderank3tadpole2} and~\eqref{eq:amplituderank3sunrise} are examples of divergent, non-melonic amplitudes.
Thus, in renormalized quantities they contribute as much as the melonic contributions.
Such effects make tensorial field theory an interesting candidate to find new regimes of the theory beyond the melonic one.

\subsection*{Acknowledgments}
I would like to thank A.~Hock and R.~Wulkenhaar for discussions on the forest formula in mat\-rix field theory, as well as the referees for very valuable comments and suggestions.
I am deeply grateful to D.~Kreimer for advice and support, in particular in developing the research project this work is part of which is
funded by the Deutsche Forschungsgemeinschaft (DFG, German Research Foundation) under the project number 418838388.
Further support comes from the DFG under Germany's Excellence Strategy EXC 2044--390685587, Mathematics M\"unster: Dynamics--Geometry--Structure.

\pdfbookmark[1]{References}{ref}
\LastPageEnding


\begin{thebibliography}{99}
\footnotesize\itemsep=0pt

\bibitem{Aldous:1991kr}
Aldous D., The continuum random tree.~{I}, \textit{Ann. Probab.} \textbf{19}
 (1991), 1--28.

\bibitem{BenGeloun:2014gp}
Ben~Geloun J., Renormalizable models in rank {$d\geq 2$} tensorial group field
 theory, \href{https://doi.org/10.1007/s00220-014-2142-6}{\textit{Comm. Math. Phys.}} \textbf{332} (2014), 117--188,
 \href{https://arxiv.org/abs/1306.1201}{arXiv:1306.1201}.

\bibitem{BenGeloun:2013fw}
Ben~Geloun J., Rivasseau V., A renormalizable 4-dimensional tensor field
 theory, \href{https://doi.org/10.1007/s00220-012-1549-1}{\textit{Comm. Math. Phys.}} \textbf{318} (2013), 69--109,
 \href{https://arxiv.org/abs/1111.4997}{arXiv:1111.4997}.

\bibitem{BenGeloun:2013dl}
Ben~Geloun J., Samary D.O., 3{D} tensor field theory: renormalization and
 one-loop {$\beta$}-functions, \href{https://doi.org/10.1007/s00023-012-0225-5}{\textit{Ann. Henri Poincar\'e}} \textbf{14}
 (2013), 1599--1642, \href{https://arxiv.org/abs/1201.0176}{arXiv:1201.0176}.

\bibitem{Bergbauer:2006vm}
Bergbauer C., Kreimer D., Hopf algebras in renormalization theory: locality and
 {D}yson--{S}chwinger equations from {H}ochschild cohomology, in Physics and
 Number Theory, \textit{IRMA Lect. Math. Theor. Phys.}, Vol.~10, \href{https://doi.org/10.4171/028-1/4}{Eur. Math.
 Soc.}, Z\"urich, 2006, 133--164, \href{https://arxiv.org/abs/hep-th/0506190}{arXiv:hep-th/0506190}.

\bibitem{Blaschke:2013kq}
Blaschke D.N., Gieres F., Heindl F., Schweda M., Wohlgenannt M., {BPHZ}
 renormalization and its application to non-commutative field theory,
 \href{https://doi.org/10.1140/epjc/s10052-013-2566-8}{\textit{Eur. Phys.~J.~C}} \textbf{73} (2013), 2566, 16~pages,
 \href{https://arxiv.org/abs/1307.4650}{arXiv:1307.4650}.

\bibitem{Bogoliubow:1957eu}
Bogoliubow N.N., Parasiuk O.S., \"Uber die {M}ultiplikation der
 {K}ausalfunktionen in der {Q}uantentheorie der {F}elder, \href{https://doi.org/10.1007/BF02392399}{\textit{Acta Math.}}
 \textbf{97} (1957), 227--266.

\bibitem{Bonzom:2012bg}
Bonzom V., Gurau R., Rivasseau V., Random tensor models in the large $N$ limit:
 uncoloring the colored tensor models, \href{https://doi.org/10.1103/PhysRevD.85.084037}{\textit{Phys. Rev.~D}} \textbf{85}
 (2012), 084037, 12~pages, \href{https://arxiv.org/abs/1202.3637}{arXiv:1202.3637}.

\bibitem{Borinsky:2018}
Borinsky M., Graphs in perturbation theory. Algebraic structure and
 asymptotics, \textit{Springer Theses}, \href{https://doi.org/10.1007/978-3-030-03541-9}{Springer}, Cham, 2018, \href{https://arxiv.org/abs/1807.02046}{arXiv:1807.02046}.

\bibitem{Branahl:2020vca}
Branahl J., Hock A., Wulkenhaar R., Blobbed topological recursion of the
 quartic Kontsevich model~I: Loop equations and conjectures,
 \href{https://arxiv.org/abs/2008.12201}{arXiv:2008.12201}.

\bibitem{Branahl:2020vc}
Branahl J., Hock A., Wulkenhaar R., Perturbative and geometric analysis of the
 quartic {K}ontsevich model, \href{https://doi.org/10.3842/SIGMA.2021.085}{\textit{SIGMA}} \textbf{17} (2021), 085, 33~pages,
 \href{https://arxiv.org/abs/2012.02622}{arXiv:2012.02622}.

\bibitem{Broadhurst:2001hw}
Broadhurst D.J., Kreimer D., Exact solutions of {D}yson--{S}chwinger equations
 for iterated one-loop integrals and propagator-coupling duality,
 \href{https://doi.org/10.1016/S0550-3213(01)00071-2}{\textit{Nuclear Phys.~B}} \textbf{600} (2001), 403--422,
 \href{https://arxiv.org/abs/hep-th/0012146}{arXiv:hep-th/0012146}.

\bibitem{Carrozza:2013uq}
Carrozza S., Tensorial methods and renormalization in group field theories,
 \textit{Springer Theses}, \href{https://doi.org/10.1007/978-3-319-05867-2}{Springer}, Cham, 2014, \href{https://arxiv.org/abs/1310.3736}{arXiv:1310.3736}.

\bibitem{Carrozza:2014bh}
Carrozza S., Oriti D., Rivasseau V., Renormalization of a {${\rm SU}(2)$}
 tensorial group field theory in three dimensions, \href{https://doi.org/10.1007/s00220-014-1928-x}{\textit{Comm. Math. Phys.}}
 \textbf{330} (2014), 581--637, \href{https://arxiv.org/abs/1303.6772}{arXiv:1303.6772}.

\bibitem{Carrozza:2014ee}
Carrozza S., Oriti D., Rivasseau V., Renormalization of tensorial group field
 theories: {A}belian {$U(1)$} models in four dimensions, \href{https://doi.org/10.1007/s00220-014-1954-8}{\textit{Comm. Math.
 Phys.}} \textbf{327} (2014), 603--641, \href{https://arxiv.org/abs/1207.6734}{arXiv:1207.6734}.

\bibitem{Connes:2000km}
Connes A., Kreimer D., Renormalization in quantum field theory and the
 {R}iemann--{H}ilbert problem. {I}.~{T}he {H}opf algebra structure of graphs
 and the main theorem, \href{https://doi.org/10.1007/s002200050779}{\textit{Comm. Math. Phys.}} \textbf{210} (2000),
 249--273, \href{https://arxiv.org/abs/hep-th/9912092}{arXiv:hep-th/9912092}.

\bibitem{Connes:2001hk}
Connes A., Kreimer D., Renormalization in quantum field theory and the
 {R}iemann--{H}ilbert problem. {II}.~{T}he {$\beta$}-function, diffeomorphisms
 and the renormalization group, \href{https://doi.org/10.1007/PL00005547}{\textit{Comm. Math. Phys.}} \textbf{216}
 (2001), 215--241, \href{https://arxiv.org/abs/hep-th/0003188}{arXiv:hep-th/0003188}.

\bibitem{Delepouve:2015hc}
Delepouve T., Gurau R., Phase transition in tensor models, \href{https://doi.org/10.1007/JHEP06(2015)178}{\textit{J.~High
 Energy Phys.}} \textbf{2015} (2015), no.~6, 178, 34~pages,
 \href{https://arxiv.org/abs/1504.05745}{arXiv:1504.05745}.

\bibitem{EbrahimiFard:2007fb}
Ebrahimi-Fard K., Gracia-Bond\'{\i}a J.M., Patras F., A {L}ie theoretic
 approach to renormalization, \href{https://doi.org/10.1007/s00220-007-0346-8}{\textit{Comm. Math. Phys.}} \textbf{276} (2007),
 519--549, \href{https://arxiv.org/abs/hep-th/0609035}{arXiv:hep-th/0609035}.

\bibitem{Freidel:2005jy}
Freidel L., Group field theory: an overview, \href{https://doi.org/10.1007/s10773-005-8894-1}{\textit{Internat.~J. Theoret.
 Phys.}} \textbf{44} (2005), 1769--1783, \href{https://arxiv.org/abs/hep-th/0505016}{arXiv:hep-th/0505016}.

\bibitem{Grosse:2020uo}
Grosse H., Hock A., Wulkenhaar R., Solution of the self-dual {$\Phi^4$}
 {QFT}-model on four-dimensional {M}oyal space, \href{https://doi.org/10.1007/jhep01(2020)081}{\textit{J.~High Energy Phys.}}
 \textbf{2020} (2020), no.~1, 081, 16~pages, \href{https://arxiv.org/abs/1908.04543}{arXiv:1908.04543}.

\bibitem{Grosse:2017ep}
Grosse H., Sako A., Wulkenhaar R., Exact solution of matricial {$\Phi_2^3$}
 quantum field theory, \href{https://doi.org/10.1016/j.nuclphysb.2017.10.010}{\textit{Nuclear Phys.~B}} \textbf{925} (2017), 319--347,
 \href{https://arxiv.org/abs/1610.00526}{arXiv:1610.00526}.

\bibitem{Grosse:2004bm}
Grosse H., Wulkenhaar R., Renormalisation of {$\phi^4$}-theory on
 non-commutative {${\mathbb R}^4$} to all orders, \href{https://doi.org/10.1007/s11005-004-5116-3}{\textit{Lett. Math. Phys.}}
 \textbf{71} (2005), 13--26, \href{https://arxiv.org/abs/hep-th/0403232}{arXiv:hep-th/0403232}.

\bibitem{Gurau:2011dw}
Gurau R., Colored group field theory, \href{https://doi.org/10.1007/s00220-011-1226-9}{\textit{Comm. Math. Phys.}} \textbf{304}
 (2011), 69--93, \href{https://arxiv.org/abs/0907.2582}{arXiv:0907.2582}.

\bibitem{Gurau:2012ek}
Gurau R., The complete {$1/N$} expansion of colored tensor models in arbitrary
 dimension, \href{https://doi.org/10.1007/s00023-011-0118-z}{\textit{Ann. Henri Poincar\'e}} \textbf{13} (2012), 399--423,
 \href{https://arxiv.org/abs/1102.5759}{arXiv:1102.5759}.

\bibitem{Gurau:2016}
Gurau R., Random tensors, Oxford University Press, Oxford, 2017.

\bibitem{Hepp:1966bx}
Hepp K., Proof of the {B}ogoliubov--{P}arasiuk theorem on renormalization,
 \href{https://doi.org/10.1007/BF01773358}{\textit{Comm. Math. Phys.}} \textbf{2} (1966), 301--326.

\bibitem{Hock:20}
Hock A., Matrix field theory, Ph.D.~Thesis, {WWU} {M}\"unster, 2020,
 \href{https://arxiv.org/abs/2005.07525}{arXiv:2005.07525}.

\bibitem{Kreimer:1998te}
Kreimer D., On the {H}opf algebra structure of perturbative quantum field
 theories, \href{https://doi.org/10.4310/ATMP.1998.v2.n2.a4}{\textit{Adv. Theor. Math. Phys.}} \textbf{2} (1998), 303--334,
 \href{https://arxiv.org/abs/q-alg/9707029}{arXiv:q-alg/9707029}.

\bibitem{Kreimer:2002br}
Kreimer D., Combinatorics of (perturbative) quantum field theory, \href{https://doi.org/10.1016/S0370-1573(01)00099-0}{\textit{Phys.
 Rep.}} \textbf{363} (2002), 387--424, \href{https://arxiv.org/abs/hep-th/0010059}{arXiv:hep-th/0010059}.

\bibitem{Kreimer:2006gg}
Kreimer D., Anatomy of a gauge theory, \href{https://doi.org/10.1016/j.aop.2006.01.004}{\textit{Ann. Physics}} \textbf{321}
 (2006), 2757--2781, \href{https://arxiv.org/abs/hep-th/0509135}{arXiv:hep-th/0509135}.

\bibitem{Kreimer:2006ft}
Kreimer D., Yeats K., An \'etude in non-linear {D}yson--{S}chwinger equations,
 \href{https://doi.org/10.1016/j.nuclphysbps.2006.09.036}{\textit{Nuclear Phys.~B Proc. Suppl.}} \textbf{160} (2006), 116--121,
 \href{https://arxiv.org/abs/hep-th/0605096}{arXiv:hep-th/0605096}.

\bibitem{Lionni:2019vc}
Lionni L., Marckert J.-F., Iterated foldings of discrete spaces and their
 limits: candidates for the role of Brownian map in higher dimensions,
 \href{https://arxiv.org/abs/1908.02259}{arXiv:1908.02259}.

\bibitem{Lionni:2017tk}
Lionni L., Th\"urigen J., Multi-critical behaviour of 4-dimensional tensor
 models up to order~6, \href{https://doi.org/10.1016/j.nuclphysb.2019.02.026}{\textit{Nuclear Phys.~B}} \textbf{941} (2019), 600--635,
 \href{https://arxiv.org/abs/1707.08931}{arXiv:1707.08931}.

\bibitem{Manchon:2004vi}
Manchon D., Hopf algebras, from basics to applications to renormalization,
 \href{https://arxiv.org/abs/math.QA/0408405}{arXiv:math.QA/0408405}.

\bibitem{Marckert:2006fs}
Marckert J.-F., Mokkadem A., Limit of normalized quadrangulations: the
 {B}rownian map, \href{https://doi.org/10.1214/009117906000000557}{\textit{Ann. Probab.}} \textbf{34} (2006), 2144--2202,
 \href{https://arxiv.org/abs/math.PR/0403398}{arXiv:math.PR/0403398}.

\bibitem{Oriti:2003uw}
Oriti D., Spin foam models of quantum spacetime, Ph.D.~Thesis, {C}ambridge
 University, 2003, \href{https://arxiv.org/abs/gr-qc/0311066}{arXiv:gr-qc/0311066}.

\bibitem{Oriti:2007qd}
Oriti D., Group field theory as the microscopic description of the quantum
 spacetime fluid: a new perspective on the continuum in quantum gravity,
 \textit{PoS Proc. Sci.} (2007), PoS(QG--Ph), 030, 38~pages,
 \href{https://arxiv.org/abs/0710.3276}{arXiv:0710.3276}.

\bibitem{Oriti:2015kv}
Oriti D., Ryan J.P., Th\"urigen J., Group field theories for all loop quantum
 gravity, \href{https://doi.org/10.1088/1367-2630/17/2/023042}{\textit{New~J. Phys.}} \textbf{17} (2015), 023042, 46~pages,
 \href{https://arxiv.org/abs/1409.3150}{arXiv:1409.3150}.

\bibitem{Panzer:2015bu}
Panzer E., Algorithms for the symbolic integration of hyperlogarithms with
 applications to {F}eynman integrals, \href{https://doi.org/10.1016/j.cpc.2014.10.019}{\textit{Comput. Phys. Comm.}}
 \textbf{188} (2015), 148--166, \href{https://arxiv.org/abs/1403.3385}{arXiv:1403.3385}.

\bibitem{Panzer:2018tg}
Panzer E., Wulkenhaar R., Lambert-{$W$} solves the noncommutative
 {$\Phi^4$}-model, \href{https://doi.org/10.1007/s00220-019-03592-4}{\textit{Comm. Math. Phys.}} \textbf{374} (2020), 1935--1961,
 \href{https://arxiv.org/abs/1807.02945}{arXiv:1807.02945}.

\bibitem{Pascalie:2019tu}
Pascalie R., A solvable tensor field theory, \href{https://doi.org/10.1007/s11005-019-01245-0}{\textit{Lett. Math. Phys.}}
 \textbf{110} (2019), 925--943, \href{https://arxiv.org/abs/1903.02907}{arXiv:1903.02907}.

\bibitem{Perez:2013uz}
Perez A., The spin foam approach to quantum gravity, \href{https://doi.org/10.12942/lrr-2013-3}{\textit{Living Rev.
 Relativ.}} \textbf{16} (2013), 3, 128~pages, \href{https://arxiv.org/abs/1205.2019}{arXiv:1205.2019}.

\bibitem{Pithis:2020gy}
Pithis A.G.A., Th\"urigen J., Phase transitions in TGFT: functional
 renormalization group in the cyclic-melonic potential approximation and
 equivalence to ${\rm O}(N)$ models, \href{https://doi.org/10.1007/jhep12(2020)159}{\textit{J.~High Energy Phys.}}
 \textbf{2020} (2020), no.~12, 159, 54~pages, \href{https://arxiv.org/abs/2009.13588}{arXiv:2009.13588}.

\bibitem{Raasakka:2013wa}
Raasakka M., Tanasa A., Combinatorial {H}opf algebra for the {B}en
 {G}eloun--{R}ivasseau tensor field theory, \textit{S\'em. Lothar. Combin.}
 \textbf{70} (2013), Art.~B70d, 29~pages, \href{https://arxiv.org/abs/1306.1022}{arXiv:1306.1022}.

\bibitem{Rivasseau:2012jk}
Rivasseau V., Quantum gravity and renormalization: the tensor track,
 \href{https://doi.org/10.1063/1.4715396}{\textit{AIP Conf. Proc.}} \textbf{2012} (2012), 18--29, \href{https://arxiv.org/abs/1112.5104}{arXiv:1112.5104}.

\bibitem{Rivasseau:2015im}
Rivasseau V., Why are tensor field theories asymptotically free?,
 \href{https://doi.org/10.1209/0295-5075/111/60011}{\textit{Europhys. Lett.}} \textbf{111} (2015), 60011, 6~pages,
 \href{https://arxiv.org/abs/1507.04190}{arXiv:1507.04190}.

\bibitem{Seidel:1973ww}
Seidel J.J., On two-graphs and {S}hult's characterization of symplectic and
 orthogonal geometries over {GF}(2), {T.H.}-Report, No.~73-WSK-02, Department of Mathematics, Technological
 University Eindhoven, Eindhoven, 1973.

\bibitem{Tanasa:2013kh}
Tanas\u{a} A., Kreimer D., Combinatorial {D}yson--{S}chwinger equations in
 noncommutative field theory, \href{https://doi.org/10.4171/JNCG/116}{\textit{J.~Noncommut. Geom.}} \textbf{7} (2013),
 255--289, \href{https://arxiv.org/abs/0907.2182}{arXiv:0907.2182}.

\bibitem{Tanasa:2007xa}
Tanas\u{a} A., Vignes-Tourneret F., Hopf algebra of non-commutative field
 theory, \href{https://doi.org/10.4171/JNCG/17}{\textit{J.~Noncommut. Geom.}} \textbf{2} (2008), 125--139,
 \href{https://arxiv.org/abs/0707.4143}{arXiv:0707.4143}.

\bibitem{Taylor:1971uv}
Taylor D.E., Some topics in the theory of finite groups, {O}xford University,
 1971.

\bibitem{Thurigen:2015}
Th\"urigen J., Discrete quantum geometries and their effective dimension, Ph.D.~Thesis, {H}umboldt-Universit\"at zu Berlin, 2015, \href{https://arxiv.org/abs/1510.08706}{arXiv:1510.08706}.

\bibitem{Thurigen:2021vy}
Th\"urigen J., Renormalization in combinatorially non-local field theories: the
 {H}opf algebra of 2-graphs, \href{https://doi.org/10.1007/s11040-021-09390-6}{\textit{Math. Phys. Anal. Geom.}} \textbf{24}
 (2021), 19, 26~pages, \href{https://arxiv.org/abs/2102.12453}{arXiv:2102.12453}.

\bibitem{Wulkenhaar:2019}
Wulkenhaar R., Quantum field theory on noncommutative spaces, in Advances in
 Noncommutative Geo\-metry, \href{https://doi.org/10.1007/978-3-030-29597-4_11}{Springer}, Cham, 2019, 607--690.

\bibitem{Zimmermann:1969up}
Zimmermann W., Convergence of {B}ogoliubov's method of renormalization in
 momentum space, \href{https://doi.org/10.1007/BF01645676}{\textit{Comm. Math. Phys.}} \textbf{15} (1969), 208--234.

\end{thebibliography}
\end{document}